\documentclass[12pt]{article}
\usepackage{epsfig}
\usepackage{psfrag}
\usepackage{latexsym}
\usepackage[DIV13]{typearea}
\usepackage{amsmath}
\usepackage{amssymb}
\usepackage{amsfonts}
\usepackage{cite}
\usepackage{bbold}
\usepackage[footnotesize]{caption2}
\usepackage{graphicx}
\usepackage[center,footnotesize,hang]{subfigure}
\usepackage{url}
\def\smu{\sigma^{\mu}}

\def\smub{{\bar\sigma}^{\mu}}

\def\snu{\sigma^{\nu}}

\def\snub{{\bar\sigma}^{\nu}}
\def\smn{\sigma^{\mu\nu}}
\def\smnb{{\bar\sigma}^{\mu\nu}}

\newcommand{\phit}{\varphi_T}

\newcommand{\phis}{\varphi_S}

\newcommand{\Uu}{\text{U(1)}}

\newcommand{\Zt}{\text{Z}_3}

\def\gs{\mathrel{
   \rlap{\raise 0.511ex \hbox{$>$}}{\lower 0.511ex \hbox{$\sim$}}}}
\def\ls{\mathrel{
   \rlap{\raise 0.511ex \hbox{$<$}}{\lower 0.511ex \hbox{$\sim$}}}}

\newcommand{\ba}{\begin{array}{c}}
\newcommand{\baz}{\begin{array}{cc}}
\newcommand{\bad}{\begin{array}{ccc}}
\newcommand{\ea}{\end{array}}

\newcommand{\be}{\beta}
\newcommand{\ga}{\gamma}

\newcommand{\om}{\omega}

\newcommand{\onep}{1^\prime}
\newcommand{\onepp}{1^{\prime\prime}}

\def\beq{\begin{equation}}
\def\eeq{\end{equation}}
\def\bea{\begin{eqnarray}}
\def\eea{\end{eqnarray}}
\def\bet{\begin{tabular}}
\def\eet{\end{tabular}}
\def\bes{\begin{subequations}\bea}
\def\ees{\eea\end{subequations}}

\def\be{\begin{equation}}
\def\ee{\end{equation}}
\def\bc{\begin{center}}
\def\ec{\end{center}}
\def\bea{\begin{eqnarray}}
\def\eea{\end{eqnarray}}
\def\dd{\displaystyle}
\def\nn{\nonumber}

\def\ov{\overline}

\catcode`@=11
\def\marginnote#1{}
\newcount\hour
\newcount\minute
\newtoks\amorpm
\hour=\time\divide\hour by60
\minute=\time{\multiply\hour by60 \global\advance\minute by-\hour}
\edef\standardtime{{\ifnum\hour<12 \global\amorpm={am}%
        \else\global\amorpm={pm}\advance\hour by-12 \fi
        \ifnum\hour=0 \hour=12 \fi
        \number\hour:\ifnum\minute<10 0\fi\number\minute\the\amorpm}}
\edef\militarytime{\number\hour:\ifnum\minute<10 0\fi\number\minute}
\def\draftlabel#1{{\@bsphack\if@filesw {\let\thepage\relax
   \xdef\@gtempa{\write\@auxout{\string
      \newlabel{#1}{{\@currentlabel}{\thepage}}}}}\@gtempa
   \if@nobreak \ifvmode\nobreak\fi\fi\fi\@esphack}
        \gdef\@eqnlabel{#1}}
\def\@eqnlabel{}
\def\@vacuum{}
\def\draftmarginnote#1{\marginpar{\raggedright\scriptsize\tt#1}}
\def\draft{\oddsidemargin 0.0truein
        \def\@oddfoot{\sl preliminary draft \hfil
        \rm\thepage\hfil\sl\today\quad\militarytime}
        \let\@evenfoot\@oddfoot \overfullrule 3pt
        \let\label=\draftlabel
        \let\marginnote=\draftmarginnote
   \def\@eqnnum{(\theequation)\rlap{\kern\marginparsep\tt\@eqnlabel}%
\global\let\@eqnlabel\@vacuum}  }
\catcode`@=12
\begin{document}
\begin{titlepage}
\vspace*{-1cm}
\phantom{hep-ph/***}
\hfill{DFPD-08/TH/09}
\vskip 2.5cm
\begin{center}
{\Large\bf Lepton Flavour Violation in\\
\vskip .3cm
Models with $A_4$ Flavour Symmetry}
\end{center}
\vskip 0.5  cm
\begin{center}
{\large Ferruccio Feruglio}~$^{a)}$\footnote{e-mail address: feruglio@pd.infn.it},
{\large Claudia Hagedorn}~$^{b)}$\footnote{e-mail address: hagedorn@mpi-hd.mpg.de},
\\
\vskip .2cm
{\large Yin Lin}~$^{a)}$\footnote{e-mail address: yin.lin@pd.infn.it} and
{\large Luca Merlo}~$^{a)}$\footnote{e-mail address: merlo@pd.infn.it}
\\
\vskip .2cm
$^{a)}$~Dipartimento di Fisica `G.~Galilei', Universit\`a di Padova
\\
INFN, Sezione di Padova, Via Marzolo~8, I-35131 Padua, Italy
\\
\vskip .1cm
$^{b)}$~
Max-Planck-Institut f\"ur Kernphysik
\\
Postfach 10 39 80, 69029 Heidelberg, Germany
\end{center}
\vskip 0.2cm
\begin{abstract}
\noindent
We analyze lepton flavour violating transitions, leptonic magnetic dipole moments (MDMs) and electric dipole moments (EDMs)
in a class of models characterized by the flavour symmetry $A_4\times Z_3\times U(1)_{FN}$, whose choice is
motivated by the approximate tri-bimaximal mixing observed in neutrino oscillations.
We construct the relevant low-energy effective Lagrangian where these effects are dominated by
dimension six operators, suppressed by the scale $M$ of new physics. All the flavour breaking effects
are universally described by the vacuum expectation values $\langle\Phi\rangle$ of a set of spurions.
We separately analyze both a supersymmetric and a general case.
While the observed discrepancy $\delta a_\mu$ in the anomalous MDM of the muon suggests
$M$ of order of a few TeV, several data require $M$ above 10 TeV, in particular
the limit on EDM of the electron.
In the general case also the present limit on $BR(\mu\to e \gamma)$ requires
$M>10$ TeV, at least. The branching ratios for
$\mu\to e \gamma$, $\tau\to \mu\gamma$ and $\tau \to\ e \gamma$ are all expected to be of the same order.
In the supersymmetric case the constraint from $\mu\to e \gamma$ is softened and it can be satisfied
by a smaller scale $M$.
In this case both the observed $\delta a_\mu$ and the current bound
on $BR(\mu\to e \gamma)$ can be satisfied, at the price of a rather small value for $|\langle\Phi\rangle|$,
of the order of a few percents, that reflects on a similar value for $\theta_{13}$.
\end{abstract}
\end{titlepage}
\setcounter{footnote}{0}
\vskip2truecm
%
%
\section{Introduction}
Aim of this paper is to analyze the predictions of a class of models of lepton masses and mixing angles
concerning processes which violate the individual lepton number, such as $\mu\to e \gamma$
and $\tau\to\mu\gamma$. We will focus on a flavour symmetry that contains the
discrete group $A_4$ and is particularly successful in reproducing the
lepton mixing angles observed in neutrino oscillations\cite{Data,Schwetz:2007my}. Such a mixing pattern is surprisingly close to the so-called tri-bimaximal
(TB) one\cite{TB}. The experimental precision on the atmospheric and solar mixing angles has considerably improved
in the last few years and now the solar mixing angle is known with a 1$\sigma$ accuracy of about one and a half degrees.
Since its first measurement the atmospheric mixing angle showed a steady tendency
to be very close to maximal, and now its experimental error is of about four to five degrees.
At present the TB scheme perfectly matches the data and it predicts a vanishing angle $\theta_{13}$ \cite{foglinove08}:
\begin{equation}
\begin{array}{lll}
\sin^2\theta_{13}<0.032 &~~ (\theta_{13}<10.3^0)~~(2\sigma) &~~ \sin^2\theta_{13}^{TB}=0\\
\sin^2\theta_{23}=0.45^{+0.16}_{-0.09} &~~ (\theta_{23}=(42.1^{+9.2}_{-5.3})^0)~~(2\sigma) &~~ \sin^2\theta_{23}^{TB}=1/2\\
\sin^2\theta_{12}=0.326^{+0.05}_{-0.04} &~~ (\theta_{12}=(34.8^{+3.0}_{-2.5})^0)~~(2\sigma) &~~ \sin^2\theta_{12}^{TB}=1/3
\label{thetatb}
\end{array}
\end{equation}
To a high degree of accuracy the TB mixing pattern can be derived by assuming that leptons transform
as certain multiplets of the discrete flavour symmetry $A_4$\cite{TBA4,af1,af2,afl,afh,linyin}, realized at a very high energy scale $\Lambda_f$.
The $A_4$ group is spontaneously broken at a scale $\langle\varphi\rangle\ll \Lambda_f$ by a set of scalar multiplets $\varphi$,
the flavons,
whose vacuum expectation values (VEVs) are subject to a special alignment. In models that have been explicitly constructed and analyzed, this alignment is not arranged by hand,
but it is the natural consequence of the dynamics of the theory\cite{af1,af2,afl,afh,linyin}.
Moreover the TB mixing pattern is expected to be modified by corrections of order $\langle\varphi\rangle/\Lambda_f\ll 1$.
When these corrections are accounted for, the mixing angle $\theta_{13}$ is no longer vanishing and becomes proportional to
$\langle\varphi\rangle/\Lambda_f$.
In these models the total lepton number is broken at a large scale $\Lambda_L$ (possibly related to $\langle\varphi\rangle$) and
light neutrinos are assumed to be of Majorana type. Depending on the specific way the total lepton number is violated
(either by higher dimensional operators or via the see-saw mechanism), the neutrino mass spectrum can have different
properties that can be tested in future experiments.

It would be highly desirable to reveal the characteristic $A_4$ symmetry pattern in other types of observables,
not directly related to neutrino properties. Such a possibility becomes realistic if there is new physics at a much lower
energy scale $M$, around $1\div 10$ TeV. 
Indeed we have several indications, both from the experimental and from the theory side,
that this can be the case. For instance, the observed discrepancy in the anomalous magnetic moment of the muon, the overwhelming evidence
of dark matter, the evolution of the gauge coupling constants towards a common high-energy value and the solution of the
hierarchy problem can all benefit from the existence of new particles around the TeV scale. In this paper
we assume that such a new scale exists and that the associated degrees of freedom do not provide new sources of baryon and/or lepton number violation.

Therefore, at least four different scales are present in our approach: the lepton number breaking scale $\Lambda_L$, the scale of flavour breaking $\Lambda_f$,
the scale introduced by the VEVs of the flavon fields $\langle\varphi\rangle$
and the new physics scale $M$. A generic hierarchy among the scales is
$M \ll \langle \varphi \rangle \ll \Lambda_f$ with $\Lambda_L$ expected to
be comparable to or smaller than $\Lambda_f$.

We will adopt an effective field theory approach, where the dominant physical effects of the new particles at low energies
can be described by local dimension six operators, suppressed by two powers of the new mass scale $M$ and explicitly conserving $B$ and $L$.
We can account for the flavour breaking effects by requiring invariance of these operators under the flavour
symmetry and by encoding the symmetry breaking effects in the flavon fields.
This approach is strictly related to, and indeed inspired by, that of minimal flavour violation (MFV)\cite{MFV,MLFV},
where the flavour group $G_f$ contains, in the lepton sector, SU(3)$_{e^c}\times$ SU(3)$_l$, and where the dimensionless
symmetry breaking parameters are the Yukawa couplings themselves. While such a choice has the advantage that it can accommodate any
pattern of lepton masses and mixing angles, it does not provide any clue about the origin of the
approximate TB pattern observed in the lepton mixing matrix $U_{PMNS}$.
For this reason, here we choose a flavour group that includes the discrete
factor $A_4$: $G_f=A_4\times Z_3\times U(1)_{FN}$,
a sort of minimal choice to properly account for both, the neutrino and the charged lepton mass spectra.
Once the choice of the group has been made,
we will explicitly build the operators of dimension six in the lepton sector. They will contribute to physical effects like the
anomalous magnetic moments (MDMs) of the charged leptons, their electric dipole moments (EDMs) and lepton flavour violating (LFV)
transitions like $\mu\to e \gamma$ and $\tau\to\mu\gamma$
\footnote{For a discussion on LFV without a specific flavour symmetry, see for instance \cite{Borzumati:1986qx,Ciuchini:2007ha,LFVwithoutFlavour,Masiero:2004js}. For a recent review on the subject see ref. \cite{Raidal}.}. 
We will separately treat the general case
where no further requirement is enforced and the supersymmetric case, where additional constraints are present.

Among the observable quantities that we have analyzed, only the anomalous MDM of the muon suggests
a relatively small scale $M$, of the order of few TeV. Apart from this indication, in the non-supersymmetric case
there are several data that require $M$ above 10 TeV. First of all the EDM of the electron that represents the
strongest bound: $M>80$ TeV. The limits on the $\tau$ decays $\tau^-\to \mu^+ e^- e^-$ and $\tau^-\to e^+ \mu^- \mu^-$
push the scale $M$ above approximately 15 TeV. Finally, to satisfy the present limit on $BR(\mu\to e \gamma)$, we
need $M>10$ TeV,  at least. We also find that the branching ratios for
$\mu\to e \gamma$, $\tau\to \mu\gamma$ and $\tau \to\ e \gamma$ are all of the same order.
Given the present limit on $BR(\mu\to e \gamma)$, this implies that $\tau\to \mu\gamma$ and $\tau \to\ e \gamma$
have rates much smaller than the present (and near future) sensitivity.
In the supersymmetric case the bound from $\mu\to e \gamma$ is softened, due to a cancellation occurring
in the relevant dipole operator. 
Depending on $|\langle\varphi\rangle/\Lambda_f|$, whose typical value lies in the range $0.001\div 0.05$,
$M$ can be as small as $0.7\div 14$ TeV. Therefore in a portion of the parameter space our model
can simultaneously fit the observed discrepancy of the anomalous MDM of the muon and respect the current bound
on $BR(\mu\to e \gamma)$. The allowed values for $\langle\varphi\rangle/\Lambda_f$ are rather small,
of the order of a few percents at most and this represents also the expected range of $\theta_{13}$.
To tolerate such a low scale $M$, additional mechanisms are required to suppress at least the electron EDM
and the rates for $\tau^-\to \mu^+ e^- e^-$ and $\tau^-\to e^+ \mu^- \mu^-$.

The paper is organized as follows:
in the second section we give a more extended overview of this approach, without specifying the flavour symmetry
and its breaking pattern. Then we move to the class of models we are interested in. We will characterize them in section 3.
We will display the corresponding low-energy effective Lagrangian in section 4.
Subsequently we will derive the predictions for leptonic MDMs, EDMs and LFV transitions both without (section 5) and with
(section 6) the additional constraints implied by supersymmetry. These sections contain the main results of our paper.
They are compared to those of MFV in section 7 and summarized in section 8. More technical aspects are discussed in 
appendix A and B.
%
%
\section{Low-energy effective Lagrangian for lepton flavour violation}

We consider the possibility that the pattern of lepton masses and mixing angles that
emerged in the last years with the discovery of neutrino oscillations is dictated by some flavour symmetry,
realized at a very high energy scale $\Lambda_f$\cite{reviewaf}.
So far all the attempts to build a theory of fermion masses based
on a flavour symmetry indicate that such a symmetry cannot be exact and that the main features of the fermion
spectrum are reproduced by specific symmetry breaking patterns. In particular, the closeness of the
observed lepton mixing to the TB mixing pattern is successfully reproduced in models
with spontaneously broken flavour symmetries and special vacuum alignment properties.
Here we assume that the flavour group $G_f$, which will be specified later on,
is spontaneously broken by a set of adimensional small parameters $\langle\Phi\rangle$, $\vert\langle\Phi\rangle\vert\ll 1$. These
should be interpreted as ratios between VEVs of flavon scalar fields $\varphi$, transforming non-trivially under $G_f$,
and some fundamental scale $\Lambda_f$, above which an ultraviolet completion of the theory describes the details
of the flavour dynamics. The consistency of this picture requires that $\vert\langle\varphi\rangle\vert\ll \Lambda_f$.
We also assume that the total lepton number (or better, the combination $B-L$) is violated at a large scale $\Lambda_L$, so that the light neutrinos
get their masses through the low-energy operator\cite{Weinberg:1979sa}:
\be
{\cal L}_\nu=\frac{1}{\Lambda_L}({\tilde H}^\dagger l)^T Y  ({\tilde H}^\dagger l)+h.c.
\ee
where $l$ denotes the three lepton doublets of the standard model (SM) and $H$ is the Higgs scalar doublet\footnote
{We adopt  two-component spinor notation, so for example
$e$ ($\bar{e}^c$) denotes the left-handed
(right-handed) component of the electron field.
For instance, in terms of the four-component spinor
$\psi^T_e = (e ~\ov{e}^c)$,
the bilinears $\ov{e} \snub e $ and
$e^c \snu \ov{e}^c$ correspond  to
$\ov{\psi}_{e} \ga^\nu P_L \psi_{e}$ and $\ov{\psi}_e \ga^\nu
P_R \psi_e $ [$P_{L,R} = \frac12 (1\mp \ga^5)$]
respectively.
We  take $\smu \equiv (1,\vec{\sigma})$, $\smub\equiv
(1,-\vec{\sigma})$, $\smn  \equiv \frac14 (\smu \snub -\snu\smub)$,
$\smnb  \equiv \frac14 (\smub \snu -\snub\smu)$ and
$g_{\mu\nu}= {\rm diag}(+1, -1, -1, -1)$, where
$\vec{\sigma} = (\sigma^1, \sigma^2, \sigma^3)$ are the
$2\times 2$ Pauli matrices. Here the four-component matrix
$\gamma^\mu$ is in the chiral basis, where the 2$\times$2 blocks along the diagonal vanish,
the upper-right block is given by $\smu$ and the lower-left block is equal to $\smub$.}.
In this scenario, the Yukawa couplings $y_f$ $(f=l,u,d)$ of the SM and the matrix $Y$ in ${\cal L}_\nu$
become functions of $\langle\Phi\rangle$:
\be
y_f=y_f\left(\langle\Phi\rangle\right)~~~,~~~~~~~Y=Y\left(\langle\Phi\rangle\right)~~~,
\label{yuk}
\ee
so that, by treating $\langle\Phi\rangle$ as spurions transforming under $G_f$ as the corresponding
parent scalar fields, the whole theory is formally invariant under $G_f$.
Under the assumption that $\vert\langle\Phi\rangle\vert\ll 1$, the functions $y_f$ and $Y$ can be expanded in powers of
$\langle\Phi\rangle$ and only a limited number of terms gives a non-negligible contribution
\footnote{An exception occurs when the dimension five operator ${\cal L}_\nu$ arises from the see-saw mechanism.
For instance, if right-handed neutrinos are present, there can be renormalizable interactions between them and the flavons.
By integrating out the right-handed neutrinos, we find that $\Lambda_L$ itself is proportional to the mass scale associated with
$\langle\varphi\rangle$. In this case the dimension five operator contains a simple pole in $\langle\Phi\rangle$ and is no longer
local in the symmetry breaking parameters. For operators violating the lepton number $L$ 
we will allow for a generalized power expansion containing
poles in $\langle\Phi\rangle$.}.

Even when $G_f$ and $\langle\Phi\rangle$ are completely specified, as in the case we are going to consider in this paper,
if $\Lambda_L$ and $\Lambda_f$ are much larger than the electroweak scale, we are left with a limited number of tests of the above picture.
Of course, there are still several parameters that remain to be measured or constrained: neutrinoless double beta decay,
the mixing angle $\theta_{13}$, the type of neutrino mass ordering, the absolute neutrino mass scale and the CP-violating phases.
However it would be highly desirable to find evidence of the flavour symmetry in other types of processes.
Such a possibility opens up if there is new physics at a much closer
energy scale $M$, around $1\div 10$ TeV. Here we assume that such a new scale exists
and that the associated degrees of freedom do not provide new sources of baryon and/or lepton number violation
\footnote{In our convention, the masses of the new particles
are of order $g M/4\pi$, if the underlying theory is weakly interacting with a typical coupling constant $g$.}.

Therefore, in the language of effective field theories, the dominant physical effects of the new particles at low energies
can be described by dimension six operators, suppressed by two powers of the new mass scale $M$ and explicitly conserving $B$ and $L$.
If we focus on the lepton sector only, the leading terms of the relevant effective Lagrangian are:
\be
{\cal L}_{eff}={\cal L}_{KT}+{e^c}^T H^\dagger y_l~ l +{\cal L}_\nu+
i\frac{e}{M^2} {e^c}^T H^\dagger \sigma^{\mu\nu} F_{\mu\nu} {\cal M}  l +h.c.+[\tt 4-fermion~~ operators]
\label{leff}
\ee
where ${\cal L}_{KT}$ stands for the kinetic terms, $e$ is the electric charge and $e^c$ the set of SU(2) lepton singlets.
The complex three by three matrix ${\cal M}$, with indices in the flavour space, is a function of $\langle\Phi\rangle$:
\be
{\cal M}={\cal M}\left(\langle\Phi\rangle\right)
\ee
Invariance under SU(2)$\times$ U(1) gauge transformations is guaranteed if $F_{\mu\nu}$ is any arbitrary
combination of $B_{\mu\nu}$, the field strength of U(1), and $\sigma^a W^a_{\mu\nu}$, the non-Abelian field strength of SU(2).
Here we are interested in the component of this combination along the direction of the unbroken U(1)$_{em}$
and we identify $F_{\mu\nu}$ with the electromagnetic field strength.
We can imagine that we derive such an effective Lagrangian from a fundamental theory by integrating out
two different sets of modes. In a first step we can integrate out the flavon fields and the possible degrees of freedom
associated with the violation of $B-L$, thus obtaining a complete set of mass terms for all the light particles including neutrinos,
charged fermions and the quanta with masses around the TeV scale. Subsequently, around the scale $M=1\div 10$ TeV, we integrate out
these additional quanta and we generate the other operators listed in eq. (\ref{leff}).
The latter are still invariant under $G_f$, once
we treat the symmetry breaking parameters as spurions.

In this way we can capture all flavour symmetry breaking effects and keep a high degree of predictability
since the expansion in the small symmetry breaking parameters can be truncated after few terms
\footnote{A non-analytic dependence on the symmetry breaking parameters $\langle \Phi \rangle$
can be induced by renormalization group running from the high energy scale $\langle \varphi \rangle$
down to energies below the electroweak scale. Such a dependence can only be determined when
all the degrees of freedom lighter than $\langle \varphi \rangle$ are known. In MFV similar effects have been
studied in ref. \cite{colangelo}.}.
Thereby, the same symmetry breaking parameters that control lepton masses and mixing angles also control
the flavour pattern of the other operators in ${\cal L}_{eff}$. Moreover the effects described by these operators
are suppressed by $1/M^2$ and not by inverse powers of the larger scales $\Lambda_f$ and $\Lambda_L$ and this opens up the possibility that
they might be observable in the future.

In a field basis where the kinetic terms are canonical and the charged lepton
mass matrix is diagonal, where we will denote vectors and matrices with a hat,
the real and imaginary parts of the matrix elements
$\hat{\cal M}_{ii}$ are proportional to the MDMs $a_i$
and to the EDMs $d_i$ of charged leptons, respectively\cite{Raidal,Brignole:2004ah,Ciuchini:2007ha,MLFV}:
\be
a_i=2 m_i \frac{v}{\sqrt{2} M^2}Re \hat{\cal M}_{ii}~~~,~~~~~~~d_i=e \frac{v}{\sqrt{2} M^2}Im \hat{\cal M}_{ii}~~~~~~~~~~(i=e,\mu,\tau)~~~,
\ee
\vskip 0.2cm
where $v\approx 246$ GeV is the electroweak breaking scale, defined as $\sqrt{2}$ times the VEV of the real, electrically neutral, component of $H$. The off-diagonal elements
$\hat{\cal M}_{ij}$ describe the amplitudes for the LFV transitions 
$\mu\to e \gamma$, $\tau\to\mu\gamma$ and $\tau\to e \gamma$\cite{Arganda:2005ji,Raidal,Ciuchini:2007ha,MLFV,Brignole:2004ah,
Borzumati:1986qx,BranchingRatios}:
\be
\frac{BR(l_i\to l_j\gamma)}{BR(l_i\to l_j\nu_i{\bar \nu_j})}=\frac{12\sqrt{2}\pi^3 \alpha}{G_F^3 m_i^2 M^4}\left(\vert\hat{\cal M}_{ij}\vert^2+\vert\hat{\cal M}_{ji}\vert^2\right)
\ee
\vskip 0.2cm
where $\alpha$ is the fine structure constant, $G_F$ is the Fermi constant and $m_i$ is the mass of the lepton $l_i$.
Finally the four-fermion operators describe other flavour violating processes
like  $\tau^-\to \mu^+ e^- e^-$ and $\tau^-\to e^+ \mu^- \mu^-$. In this paper our focus will be mainly on the processes
$\mu\to e \gamma$ and $\tau\to\mu\gamma$ and we will make only some comments on the four-fermion operators.
%
%
\section{Models with $A_4$ flavour symmetry}

So far the discussion has been rather general and applies to any theory with a spontaneously broken flavour symmetry.
Several models have been proposed to produce the TB mixing scheme\cite{TBA4,af1,af2,afl,afh,linyin,TBother} and among the most economic and simplest ones are those based on the
discrete group $A_4$\cite{TBA4,af1,af2,afl,afh,linyin}. $A_4$ is the group of even permutations of four objects, isomorphic to the group of discrete rotations in the
three-dimensional space that leave invariant a regular tetrahedron. It is generated by two elements
$S$ and $T$ obeying the relations\cite{A4Presentation}:
\be
S^2=(ST)^3=T^3=1~~~.
\label{$A_4$}
\ee
It has three independent one-dimensional representations, $1$, $1'$ and $1''$ and one three-dimensional representation $3$.
We present a set of generators $S$ and $T$ for the various representations, and the relevant multiplication
rules in appendix A. The group $A_4$ has two obvious subgroups: $G_S$, which is a reflection subgroup
generated by $S$, and $G_T$, which is the group generated by $T$, isomorphic to $Z_3$. These subgroups are of interest
for us because $G_S$ and $G_T$ are the relevant low-energy symmetries
of the neutrino and the charged-lepton sectors at leading order, respectively. The TB mixing is then a direct consequence of this
special symmetry breaking pattern, which is achieved via the vacuum misalignment of triplet scalar fields.
If $\varphi=(\varphi_1,\varphi_2,\varphi_3)$ denotes the generic scalar triplet, the VEV
\be
\langle\varphi\rangle \propto (1,1,1)
\label{unotre}
\ee
breaks $A_4$ down to $G_S$, while
\be
\langle\varphi\rangle \propto (1,0,0)
\label{unozero}
\ee
breaks $A_4$ down to $G_T$.

Concerning the flavour group $G_f$, following ref. \cite{af1,af2,afl} we will choose
\be
G_f=A_4\times Z_3\times U(1)_{FN}
\label{flavgroup}
\ee
The three factors in $G_f$ play different roles. The spontaneous breaking of the first one, $A_4$, is directly responsible for the TB mixing.
The $Z_3$ factor is a discrete version of the total lepton number and is needed in order to avoid large mixing effects
between the flavons that give masses to the charged leptons and those giving masses to neutrinos. Finally, $U(1)_{FN}$ is
responsible for the hierarchy among charged fermion masses\cite{fn}.
The flavour symmetry breaking sector of the model includes the scalar fields $\varphi_T$, $\varphi_S$, $\xi$ and $\theta$. The transformation properties of the lepton
fields $l$, $e^c$, $\mu^c$, $\tau^c$, of the electroweak scalar doublet $H$ and of the flavon fields are summarized in table 1.
We also assume the following pattern of VEVs for the flavon fields:
\bea
\frac{\langle\varphi_T\rangle}{\Lambda_f}&=&(u,0,0)+(c_1 u^2,c_2 u^2,c_3 u^2)+O(u^3)\nonumber\\
\frac{\langle\varphi_S\rangle}{\Lambda_f}&=& c_b(u,u,u)+O(u^2)\nonumber\\
\label{vevs}
\frac{\langle\xi\rangle}{\Lambda_f}&=&c_a u+O(u^2)\\
\frac{\langle\theta\rangle}{\Lambda_f}&=&t\nn
\eea
where $c_{1,2,3,a,b}$ are numbers of order one, while $u$ and $t$ are the small symmetry breaking parameters of the theory.
All these quantities are in general complex.
Notice that we have explicitly displayed the sub-leading $O(u^2)$ corrections only for the field $\varphi_T$.
Those for the fields $\varphi_S$ and $\xi$ are not needed in the present analysis: they have no special pattern and are relevant
when discussing corrections to neutrino masses and mixing angles.
It has been shown in ref. \cite{af2} how it is possible to achieve this pattern in a natural way, as the result
of the minimization of the scalar potential of the theory.
In that case the model was supersymmetric, but an analogous pattern can also be obtained
in other, non-supersymmetric, versions. Note that in the supersymmetric case the $O(u^2)$ corrections to $\varphi_T$ obey $c_2=c_3$.
\begin{table}[!ht]
\centering
                \begin{math}
                \begin{array}{|c||c|c|c|c||c|c|c|c|c|}
                    \hline
                    &&&&&&&&& \\[-9pt]
                    \text{Field} & l & e^c & \mu^c & \tau^c & H & \phit & \phis & \xi & \theta \\[10pt]
                    \hline
                    &&&&&&&&&\\[-9pt]
                    A_4 & 3 & 1 & \onepp & \onep & 1 & 3 & 3 & 1 &  1 \\[3pt]
                    \hline
                    &&&&&&&&&\\[-9pt]
                    \Zt & \om & \om^2 & \om^2 & \om^2  & 1 & 1& \om & \om  & 1 \\[3pt]
                    \hline
                    &&&&&&&&&\\[-9pt]
                    \Uu_{FN} & 0 & 2 & 1 & 0  & 0 & 0 & 0 & 0 & -1  \\[3pt]
                    \hline
                \end{array}
               \end{math}
            \caption{ The transformation rules of the fields under the
            symmetries associated with the groups \rm{$A_4$}, \rm{$\Zt$} and
            \rm{$\Uu_{FN}$}.}
            \end{table}
\vskip 0.2cm
\noindent
The flavour group of eq. (\ref{flavgroup}) and the set of flavon fields and their VEVs of eq. (\ref{vevs})
can be used to obtain neutrino masses with an approximate TB mixing, both in the presence of right-handed neutrinos
from the see-saw mechanism and directly from a set of higher dimensional operators\cite{af2} (with the only difference that, in the first case, the flavons $\varphi_S$ and $\xi$
transform with $\omega^2$ under $Z_3$). Therefore the effective Lagrangian, (\ref{leff}), and the discussion of the related
LFV processes, is exactly the same in both cases, at variance with MFV where the
flavour group in the see-saw case becomes larger\cite{MLFV}.
\vskip 0.2cm
Before going into the details of the model, we recall that at the leading order the Yukawa couplings in eq. (\ref{yuk})
are given by:
\be
y_l=\left(
\begin{array}{ccc}
\hat{y}_e t^2 & 0& 0\\
0& \hat{y}_\mu t& 0\\
0& 0& \hat{y}_\tau
\end{array}
\right) u~~~,
\label{yf}
\ee
\be
Y=\left(
\begin{array}{ccc}
a+2 b/3& -b/3& -b/3\\
-b/3& 2b/3& a-b/3\\
-b/3& a-b/3& 2 b/3
\end{array}
\right) u~~~,
\label{Y}
\ee
where $\hat{y}_e$, $\hat{y}_\mu$, $\hat{y}_\tau$, $a$ and $b$ are numbers of order one.
At this order the mass matrix for
the charged leptons is diagonal with the relative hierarchy described by the parameter $t$.
We will take
\be
\vert t\vert\approx 0.05~~~.
\label{tbound}
\ee
In the same approximation,
the neutrino mass matrix is diagonalized by the transformation:
\be
U_{TB}^T m_\nu U_{TB} =\frac{v^2}{\Lambda_L}{\tt diag}(a+b,a,-a+b) u~~~,
\ee
where $U_{TB}$, up to phases, is the unitary matrix of the TB mixing\cite{TB}:
\be
U_{TB}=\left(
\begin{array}{ccc}
\sqrt{2/3}& 1/\sqrt{3}& 0\\
-1/\sqrt{6}& 1/\sqrt{3}& -1/\sqrt{2}\\
-1/\sqrt{6}& 1/\sqrt{3}& +1/\sqrt{2}
\end{array}
\right)~~~.
\label{UTB}
\ee
The symmetry breaking parameter $u$ should lie in the range
\be
0.001<\vert u\vert < 0.05~~~,
\label{ubound}
\ee
the lower bound coming from the requirement that the Yukawa coupling of the $\tau$ does not exceed $4 \pi$, and the upper bound coming from the
requirement that the higher order corrections, so far neglected, do not modify too much the leading TB mixing.
Indeed, the inclusion of higher order corrections modifies all mixing angles by quantities of relative order $u$, especially we should keep the agreement between the
predicted and measured value of the solar angle within few degrees. The unknown angle $\theta_{13}$ is expected to be of order $|u|$.
Notice that, in most of the allowed range, we have $|u|<|t|$. Such a framework can also be extended to the quark sector\cite{afh,QuarkExtention}.
\vskip 0.5 cm
%
%
\section{Low-energy effective Lagrangian with $A_4$ flavour symmetry}
\vskip 0.5 cm
To evaluate the matrix of dipole moments ${\cal M}$ we should analyze the Lagrangian of the model:
\be
{\cal L}_{eff}={\cal L}_{KT}+{\cal L}_Y+{\cal L}_{dip}+...
\label{leel}
\ee
Each of the terms on the right-hand side can be thought of as an expansion in powers of the flavon fields. Since we have two independent
symmetry breaking parameters, $u$ and $t$, see eqs. (\ref{vevs}), we consider a double expansion of ${\cal L}_{eff}$ in powers of $u$ and $t$.
In this expansion we keep terms up to the second order in $u$, i.e. terms quadratic in the fields $\varphi_{S,T}$ and $\xi$.
The expansion in the parameter $t$, responsible for the breaking of the Froggatt-Nielsen U(1)$_{FN}$ symmetry, will be stopped
at the first non-trivial order, that is by allowing as many powers of the $\theta$ field as needed in order to
obtain non-vanishing values for all the entries of the matrices describing lepton masses and dipole moments
\footnote
{Concerning the kinetic terms we observe that we can additionally write down operators involving the total invariant
$\theta^{\dagger}\theta=|\theta|^{2}$. These contribute to the diagonal elements of
the kinetic terms. In the kinetic terms of the left-handed fields they can be safely neglected, since the
leading order correction is of $O(u)$. For the kinetic terms of the right-handed fields, they contribute at the
same order as the terms arising through a double flavon insertion.}.
Finally, second order corrections in $u$ also arise from the sub-leading terms of the VEV $\langle\varphi_T\rangle$, eq. (\ref{vevs}),
and will be included in our estimates.
\subsection{Kinetic terms}
The expansion of the kinetic terms can be written as:
\be
{\cal L}_{KT}={\cal L}_{KT}^{(0)}+{\cal L}_{KT}^{(1)}+{\cal L}_{KT}^{(2)}+...
\ee
The leading order term is given by
\be
{\cal L}_{KT}^{(0)}=i k_0\sum_{i=1}^3\bar{l}_i \bar{\sigma}^\mu D_\mu l_i+i \sum_{i=1}^3 (k_0^c)_i \bar{l^c}_i \bar{\sigma}^\mu D_\mu l^c_i
\ee
where the quantities $k_0$ and $(k_0^c)_i$ can be further expanded in powers of the field $\theta$:
\be
k_0=1+{\hat k}\dd\frac{\vert\theta\vert^2}{\Lambda_f^2}+...~~~,~~~~~~~~~~
(k_0^c)_i=1+{\hat k}^c_i\dd\frac{\vert\theta\vert^2}{\Lambda_f^2}+...
\ee
Note that for our purposes it is enough to expand $k_0$ up to the zeroth order and $(k_0^c)_i$ up to the second order in $t$.
At the first order in $u$ we have only one type of operator contributing to the kinetic terms
\be
{\cal L}_{KT}^{(1)}=i \dd\frac{k}{\Lambda_f}  (\varphi_T \bar{l} \bar{\sigma}^\mu D_\mu l)+h.c.
\ee
where $k$ is a coefficient of order one. Here and in what follows, by a parenthesis $(\cdot\cdot\cdot)$ we denote an
invariant under $A_4$, $(\cdot\cdot\cdot)'$ denotes the $1'$ singlet and so on.
At the second order in $u$ we have a richer structure:
\be
{\cal L}_{KT}^{(2)}={\cal L}_{KTL}^{(2)}+{\cal L}_{KTR}^{(2)}~~~,
\ee
with the labels $L$ and $R$ referring to lepton doublets $l$ and singlets $e^c$, $\mu^c$, $\tau^c$, respectively.
For lepton doublets we find:
\be
{\cal L}_{KTL}^{(2)}=i \sum_{i=1}^7 \dd\frac{k_i}{\Lambda_f^2}  (X_i \bar{l} \bar{\sigma}^\mu D_\mu l)
\label{here}
\ee
where $X$ is the list of $Z_3$-invariant operators, bilinear in the flavon fields $\varphi_{S,T}$ and $\xi$,
\be
X=\left\{\xi^\dagger\xi,\varphi_T^2,(\varphi_T^\dagger)^2,\varphi_T^\dagger\varphi_T,\varphi_S^\dagger\varphi_S,\xi^\dagger\varphi_S,\varphi_S^\dagger\xi\right\}~~~,
\label{Xlist}
\ee
and there are obvious relations among the coefficients $k_i$  in order to assure hermiticity.
Note that the sum in eq. (\ref{here}) runs over all bilinears which can couple to form $A_4$-invariants. Whether
they lead to a non-trivial contribution to the sum depends on the VEVs of the flavons.
For lepton singlets, we can distinguish a diagonal contribution and a non-diagonal one:
\be
{\cal L}_{KTR}^{(2)}=[{\cal L}_{KTR}^{(2)}]_{d}+[{\cal L}_{KTR}^{(2)}]_{nd}
\ee
\be
[{\cal L}_{KTR}^{(2)}]_d=i \dd\frac{1}{\Lambda_f^2}\sum_{i=1}^5 \left[ (k_e^c)_i (X_i)\bar{e^c} \bar{\sigma}^\mu D_\mu e^c+(k_\mu^c)_i(X_i)\bar{\mu^c} \bar{\sigma}^\mu D_\mu \mu^c+(k_\tau^c)_i(X_i) \bar{\tau^c} \bar{\sigma}^\mu D_\mu \tau^c\right]
\ee
\bea
[{\cal L}_{KTR}^{(2)}]_{nd}&=&i \sum_{i=2}^5 \left[ \dd\frac{(k_{e\mu}^c)_i}{\Lambda_f^3} (X_i)' \theta^\dagger  \bar{e^c} \bar{\sigma}^\mu D_\mu \mu^c+h.c.\right]\nn\\
\label{here2}
&+&i \sum_{i=2}^5 \left[\dd\frac{(k_{e\tau}^c)_i}{\Lambda_f^4} (X_i)'' (\theta^\dagger)^2  \bar{e^c} \bar{\sigma}^\mu D_\mu \tau^c  +h.c.\right]\\
&+& i \sum_{i=2}^5 \left[\dd\frac{(k_{\mu\tau}^c)_i}{\Lambda_f^3} (X_i)' \theta^\dagger  \bar{\mu^c} \bar{\sigma}^\mu D_\mu \tau^c+h.c.\right]\nn
\eea
Due to the structure of the flavon VEVs only the term with $i=5$ gives a non-vanishing
        contribution in the sum in eq. (\ref{here2}).
When the flavon fields acquire a VEV according to the pattern assumed in eq. (\ref{vevs}), the term ${\cal L}_{KT}$ gives rise to non-canonical kinetic
terms for the lepton fields, of the following form:
\be
{\cal L}_{KT}=i K_{ij}\bar{l}_i \bar{\sigma}^\mu D_\mu l_j+i K^c_{ij}\bar{l^c}_i \bar{\sigma}^\mu D_\mu l^c_j
\ee
\vskip 0.2 cm
\be
K=
\left(
\begin{array}{ccc}
1+O(u)& O(u^2)& O(u^2)\\
O(u^2)& 1+O(u)& O(u^2)\\
O(u^2)& O(u^2)& 1+O(u)
\end{array}
\right)
\label{K}
\ee
\vskip 0.5 cm
\be
K^c=
\left(
\begin{array}{ccc}
1+O(u^2,t^2)& O(u^2 t)& O(u^2 t^{2})\\
O(u^2 t)& 1+O(u^2,t^2)& O(u^2 t)\\
O(u^2 t^{2})& O(u^2 t)& 1+O(u^2,t^2)
\end{array}
\right)
\label{Kc}
\ee
\vskip 0.5 cm
As one can see, the
corrections to the kinetic terms which render these non-canonical are small (at most at order
$O(u)$ and $O(t^2)$). They have to be taken into account when calculating the lepton masses and the dipole transitions
induced by the matrix $\hat{\mathcal{M}}$. However, an explicit calculation shows that
at the leading order in our expansion parameters the charged lepton masses and the elements of $\hat{\cal M}$ are not affected
by these non-canonical kinetic terms\cite{Kahler}. For this reason we have omitted in $K$ and $K^c$ the order one parameters that multiply the powers of $t$ and $u$.
\subsection{Lepton mass terms}
We now move to the operators that give rise to the Yukawa couplings:
\be
{\cal L}_Y={\cal L}_l+{\cal L}_\nu
\ee
There is a part responsible for the charged lepton masses
\be
{\cal L}_l={\cal L}_l^{(1)}+{\cal L}_l^{(2)}+...
\ee
\bea
{\cal L}_l^{(1)}&=&\dd\frac{y_e}{\Lambda_f^3} \theta^2e^c H^\dagger \left(\varphi_T l\right)
+\dd\frac{y_\mu}{\Lambda_f^2} \theta\mu^c H^\dagger \left(\varphi_T l\right)'
+\dd\frac{y_\tau}{\Lambda_f} \tau^c H^\dagger \left(\varphi_T l\right)''\nonumber\\
&+&\dd\frac{y_e'}{\Lambda_f^3} \theta^2e^c H^\dagger \left(\varphi_T^\dagger l\right)
+\dd\frac{y_\mu'}{\Lambda_f^2} \theta\mu^c H^\dagger \left(\varphi_T^\dagger l\right)'
+\dd\frac{y_\tau'}{\Lambda_f} \tau^c H^\dagger \left(\varphi_T^\dagger l\right)''+h.c.
\label{yl1}
\eea
The terms of eqs. (\ref{yl1}) give rise to $y_l$ as shown in eq. (\ref{yf}), if we define
\begin{equation}
u=|u| \, \mathrm{e}^{i \psi}~~~,~~~~~~~\hat{y}_{f} = y_{f} + y_{f} ^{\prime} \mathrm{e}^{-2 i \psi} ~~~~~~~~~~(f=e,\mu,\tau)~~.
\end{equation}
At the next order in $u$ we find:
\be
{\cal L}_l^{(2)}=\sum_{i=2}^7\dd\frac{y_e^i}{\Lambda_f^4} \theta^2e^c H^\dagger \left(X_i l\right)
+\sum_{i=2}^7\dd\frac{y_\mu^i}{\Lambda_f^3} \theta\mu^c H^\dagger \left(X_i l\right)'
+\sum_{i=2}^7\dd\frac{y_\tau^i}{\Lambda_f^2} \tau^c H^\dagger \left(X_i l\right)''+h.c.
\label{yl2}
\ee
where the coefficients $y_e^i$, $y_\mu^i$ and $y_\tau^i$ are numbers of order one
\footnote{Due to assumed pattern of VEVs, eq. (\ref{vevs}), each bilinear $X_i$ gives rise
to a single contribution to the lepton masses: out of the several possible ways
of combining $X_i$ with $l$ to form the desired $A_4$ singlet, only one has a non-vanishing
VEV, at order $u^2$. Thus there is no ambiguity in the parameters $y_f^i$, when discussing the lepton mass
matrix.}.
Note that due to the special vacuum structure, eq. (\ref{vevs}), the term with $i=5$ in the sum in eq. (\ref{yl2}) gives a vanishing 
contribution. From eqs. (\ref{yl1}) and (\ref{yl2}),
after the breaking of the flavour and the electroweak symmetries, we find the following matrix of Yukawa couplings for
the charged leptons:
\begin{equation}
\label{yl_subleading}
y_{l} = \left( \begin{array}{ccc}
        \hat{y}_{e} t^2 u + z_{e}^{(1)} t^2 u^2 & z_{e}^{(2)} t^2 u^2 &  z_{e}^{(3)} t^2 u^2\\
        z_{\mu}^{(3)} t u^2 & \hat{y}_{\mu} t u + z_{\mu}^{(1)} t u^2 &  z_{\mu}^{(2)} t u^2\\
        z_{\tau}^{(2)} u^2 & z_{\tau}^{(3)} u^2 & \hat{y}_{\tau} u + z_{\tau} ^{(1)} u^2
\end{array}
\right)
\end{equation}
with
\begin{eqnarray} \nonumber
&& z_{f}^{(1)} = y_{f} c_1 +y_{f}^{\prime} c_1^{\star} \mathrm{e}^{-4 i \psi} + y_{f}^{2} +
y_{f}^{3} \mathrm{e}^{-4 i \psi} + y_{f}^{4} \mathrm{e}^{-2 i \psi}
+ y_{f}^{6} c_a^{\star} c_b \mathrm{e}^{-2 i \psi} +
y_{f}^{7} c_{a} c_{b}^{\star} \mathrm{e}^{-2 i \psi} ~~, \\ \nonumber
&& z_{f}^{(2)} =  y_{f} c_3 + y_{f}^{\prime} c_2^{\star} \mathrm{e}^{-4 i \psi}
+ y_{f}^{6} c_a^{\star} c_b \mathrm{e}^{-2 i \psi} + y_{f}^{7} c_{a} c_{b}^{\star} \mathrm{e}^{-2 i \psi} ~~,
\\ \nonumber
&& z_{f}^{(3)} =  y_f c_2 + y_f^{\prime} c_3^{\star} \mathrm{e}^{-4 i \psi}
+ y_{f}^{6} c_a ^{\star} c_b \mathrm{e}^{-2 i \psi}+ y_{f}^{7} c_{a} c_{b}^{\star} \mathrm{e}^{-2 i \psi} ~~
\end{eqnarray}
for $f=e,\mu,\tau$. We indicate complex conjugation via ${}^{\star}$ and recognize that
for $u=|u| \, \mathrm{e}^{i \psi}$ $u^{\star}$ reads $u \, \mathrm{e}^{-2 i \psi}$.
Note that the off-diagonal elements of $y_l$, proportional to $z^{(2,3)}_f$,
originate either from the sub-leading contributions to the VEV of the $\varphi_T$ multiplet
(coefficients $c_{2,3}$), or from a double flavon insertion of the type $\xi^\dagger \varphi_S$ or
$\xi \varphi_S^\dagger$ (coefficients $c_{a,b}$). Note further that in the case
$c_2=c_3$ $z_f ^{(2)}$ also equals $z_f ^{(3)}$.

Similarly, the Lagrangian giving rise to neutrino
masses can be expanded as:
\be
{\cal L}_\nu={\cal L}_\nu^{(1)}+{\cal L}_\nu^{(2)}+...
\ee
with the leading order terms:
\be
{\cal L}_\nu^{(1)}=
\dd\frac{x_a}{\Lambda_f\Lambda_L} \xi ({\tilde H}^\dagger l {\tilde H}^\dagger l) +\dd\frac{x_b}{\Lambda_f\Lambda_L} (\varphi_S {\tilde H}^\dagger l {\tilde H}^\dagger l)~~~.
\label{here3}
\ee
Due to the non-vanishing $Z_3$ charge of $\xi$ and $\varphi_{S}$ the terms equivalent to those in
eq. (\ref{here3}) with $\xi^{\dagger}$ and $\varphi_{S}^{\dagger}$ are not invariant. ${\cal L}_{\nu}^{(1)}$
leads to eq. (\ref{Y}) with $a=x_a c_a$ and $b=x_b c_b$. ${\cal L}_{\nu} ^{(2)}$ is not relevant for our discussion
here, however gives rise to deviations of relative order $|u|$ from TB mixing.

\vskip 0.2 cm
\subsection{Dipole moments}
Coming to the last term of eq. (\ref{leel}), the one contributing to the dipole moments, we observe that ${\cal L}_{dip}$ and ${\cal L}_l$  have the same structure
in flavour space and they only differ by the insertion of the electromagnetic field strength $\sigma\cdot F\equiv \sigma^{\mu\nu} F_{\mu\nu}$:
\be
{\cal L}_{dip}={\cal L}_{dip}^{(1)}+{\cal L}_{dip}^{(2)}+...
\ee
with
\bea
\label{LOM}
{\cal L}_{dip}^{(1)}&=&i\dd\frac{e}{M^2}\left[\dd\frac{\beta_e}{\Lambda_f^3} \theta^2e^c H^\dagger \sigma\cdot F \left(\varphi_T l\right)
+\dd\frac{\beta_\mu}{\Lambda_f^2} \theta\mu^c H^\dagger \sigma\cdot  F \left(\varphi_T l\right)'
+\dd\frac{\beta_\tau}{\Lambda_f} \tau^c H^\dagger \sigma\cdot F \left(\varphi_T l\right)''\right.\\
&+&\left.\dd\frac{\beta_e'}{\Lambda_f^3} \theta^2e^c H^\dagger \sigma\cdot F \left(\varphi_T^\dagger l\right)
+\dd\frac{\beta_\mu'}{\Lambda_f^2} \theta\mu^c H^\dagger \sigma\cdot  F \left(\varphi_T^\dagger l\right)'
+\dd\frac{\beta_\tau'}{\Lambda_f} \tau^c H^\dagger \sigma\cdot F \left(\varphi_T^\dagger l\right)''\right]+h.c.\nonumber
\eea
\vskip 0.1 cm
and
\be
{\cal L}_{dip}^{(2)}=i\dd\frac{e}{M^2}\sum_{i=2}^7\left[\dd\frac{\beta_e^i}{\Lambda_f^4} \theta^2e^c H^\dagger \sigma\cdot F \left(X_i l\right)
+\dd\frac{\beta_\mu^i}{\Lambda_f^3} \theta\mu^c H^\dagger \sigma\cdot  F \left(X_i l\right)'
+\dd\frac{\beta_\tau^i}{\Lambda_f^2} \tau^c H^\dagger \sigma\cdot F \left(X_i l\right)''\right]+h.c.
\label{NLOM}
\ee
\vskip 0.2 cm
Since the Yukawa and the dipole couplings involve the same fermion fields and flavons,
        all statements made concerning eqs. (\ref{yl1}) and (\ref{yl2}) also hold for eqs. (\ref{LOM}) and (\ref{NLOM}). From eqs. (\ref{LOM}) and (\ref{NLOM}),
after the breaking of the flavour and electroweak symmetries, we find the following matrix for the dipole moments:
\begin{equation}
\label{M_subleading}
\mathcal{M} = \left( \begin{array}{ccc}
        \hat{\beta}_{e} t^2 u + \gamma_{e}^{(1)} t^2 u^2 & \gamma_{e}^{(2)} t^2 u^2 &  \gamma_{e}^{(3)} t^2 u^2\\
        \gamma_{\mu}^{(3)} t u^2 & \hat{\beta}_{\mu} t u + \gamma_{\mu}^{(1)} t u^2 &  \gamma_{\mu}^{(2)} t u^2\\
        \gamma_{\tau}^{(2)} u^2 & \gamma_{\tau}^{(3)} u^2 & \hat{\beta}_{\tau} u + \gamma_{\tau} ^{(1)} u^2
\end{array}
\right)
\end{equation}
with
\begin{eqnarray} \nonumber
&& \hat{\beta}_{f} = \beta_{f} + \beta_{f} ^{\prime} \mathrm{e}^{-2 i \psi} ~~, \\ \nonumber
&& \gamma_{f}^{(1)} = \beta_{f} c_1 +\beta_{f}^{\prime} c_1^{\star} \mathrm{e}^{-4 i \psi} + \beta_{f}^{2} +
\beta_{f}^{3} \mathrm{e}^{-4 i \psi} + \beta_{f}^{4} \mathrm{e}^{-2 i \psi}
+ \beta_{f}^{6} c_a^{\star} c_b \mathrm{e}^{-2 i \psi} +
\beta_{f}^{7} c_{a} c_{b}^{\star} \mathrm{e}^{-2 i \psi} ~~, \\
&& \gamma_{f}^{(2)} =  \beta_{f} c_3 + \beta_{f}^{\prime} c_2^{\star} \mathrm{e}^{-4 i \psi}
+ \beta_{f}^{6} c_a^{\star} c_b \mathrm{e}^{-2 i \psi} + \beta_{f}^{7} c_{a} c_{b}^{\star} \mathrm{e}^{-2 i \psi} ~~,
\\ \nonumber
&& \gamma_{f}^{(3)} =  \beta_f c_2 + \beta_f^{\prime} c_3^{\star} \mathrm{e}^{-4 i \psi}
+ \beta_{f}^{6} c_a ^{\star} c_b \mathrm{e}^{-2 i \psi}+ \beta_{f}^{7} c_{a} c_{b}^{\star} \mathrm{e}^{-2 i \psi} ~~
\end{eqnarray}
for $f=e,\mu,\tau$. We observe that, in analogy to what occurs in $y_l$, the off-diagonal elements of ${\cal M}$, proportional to $\gamma^{(2,3)}_f$,
originate either from the sub-leading contributions to the VEV of the $\varphi_T$ multiplet
(coefficients $c_{2,3}$), or from a double flavon insertion of the type $\xi^\dagger \varphi_S$ or
$\xi \varphi_S^\dagger$ (coefficients $c_{a,b}$). For $c_2=c_3$ $\gamma_{f}^{(2)}$ and $\gamma_{f}^{(3)}$
become equal. This is similar to $z_f^{(2)}=z_{f}^{(3)}$ in the case of $y_l$.
\vskip 0.2cm
\subsection{Four-fermion operators}
A complete classification of SU(2)$\times$U(1), Lorentz and $G_f$ invariant four-lepton operators goes beyond the scope of this work
and here we limit ourselves to some remarks.
At variance with the operators controlled by the matrix ${\cal M}$, that vanish when either the Higgs doublet
or the flavon fields are set to zero, four-fermion operators exist that do neither require the insertion
of the Higgs field, nor the insertion of a flavon field\cite{Raidal,Arganda:2005ji,Brignole:2004ah,4FermionOperator}. A first group of terms includes
\be
\frac{1}{M^2}\ov{f^c} f^c ~\ov{{f'}^c} {f'}^c~~~,~~~~~~~~~~\frac{1}{M^2}\ov{f^c} f^c ~(\bar{l} l)~~~~~~~~~~~~~(f,f'=e,\mu,\tau)~~~.
\ee
These operators conserve the individual lepton numbers.
When they contain electrons they are bound by LEPII data, that typically require a scale $M$ above several TeV.
A second group consists of:
\be
\frac{	1}{M^2}\ov{e^c} \mu^c ~\ov{{\tau}^c} {\mu}^c
~~~,~~~~~~~~~~\frac{1}{M^2}(\bar{l} l~\bar{l} l)~~~
\label{ffop}
\ee
and their conjugates.
The second combination in eq. (\ref{ffop}) stands for several independent invariants, obtained through different contractions
of the involved $A_4$ triplets.
This other set gives rise to interactions that violate the individual lepton numbers $L_i$, with the selection rule
$\Delta L_e \Delta L_\mu \Delta L_\tau=\pm2$. For instance, one such interaction is given by
\be
({\bar l} l)'({\bar l} l)''=
{\bar l}_e l_e {\bar l}_\mu l_\mu+
{\bar l}_\mu l_\mu {\bar l}_\tau l_\tau+
{\bar l}_\tau l_\tau {\bar l}_e l_e+
\left[{\bar l}_e l_\tau {\bar l}_\mu l_\tau+
{\bar l}_\mu l_e {\bar l}_\tau l_e+
{\bar l}_\tau l_\mu {\bar l}_e l_\mu+h.c.\right]~~~.
\ee
These operators can contribute to LFV decays like $\tau^-\to \mu^+ e^- e^-$,
$\tau^-\to e^+ \mu^- \mu^-$ and their conjugate, whose branching ratios have upper bounds of the order of $10^{-7}$\cite{PDG}.
By assuming that these operators enter the low-energy Lagrangian with generic order one coefficients,
through a rough dimensional estimate we find
\be
\dd\frac{1}{G_F^2 M^4}<10^{-7} 
\ee
and we get a lower bound on the scale $M$ of the order of $15$ TeV.
A more detailed analysis of these operators and of the associated phenomenology will be presented elsewhere.
\vskip 0.5 cm
%
%
\section{Dipole transitions}
The physical observable quantities we are interested in, MDMs, EDMs and flavour violating transitions for leptons,
can be computed from the effective Lagrangian of eq. (\ref{leff}), whose terms have been detailed in the previous section.
If we work at the lowest order in the symmetry breaking parameter $u$, the kinetic terms are canonical (zeroth order
in $u$) and $y_l$ is diagonal (first order in $u$). In this approximation also the dipole matrix $\mathcal{M}$
is diagonal and it only contributes to lepton MDMs and EDMs, which we will discuss below.
In order to find also the size of the flavour violating processes we need to
include the sub-leading effects originating from insertions of the flavon fields $\varphi_{S,T}$
and $\xi$, shifts of the VEV of $\varphi_T$ and (additional) insertions of $\theta$ in the Lagrangian.
As shown above, these generate non-canonically normalized $K$
and $K^{c}$,  eq. (\ref{K}) and eq. (\ref{Kc}), and render $y_l$ and $\mathcal{M}$
non-diagonal,  eq. (\ref{yl_subleading}) and eq. (\ref{M_subleading}).
Therefore, we first have to move to the basis where the kinetic terms are canonically normalized
and the charged lepton mass matrix is diagonal. After performing the necessary steps which are outlined in appendix B,
we arrive at the following leading order result for the dipole matrix that, in this basis, will be denoted by $\mathcal{\hat M}$:
\begin{align}
\mathcal{\hat M}_{ee}&= \hat{\beta}_{e} \, t^2 \, u ~~,\nonumber\\
\mathcal{\hat M}_{e\mu}&= \frac{1}{\hat{y}_{\mu}^2} \, (z_\mu^{(3)} \hat{y}_{\mu} \hat{\beta}_{e} + \hat{y}_{\mu}^2 \gamma_{e}^{(2)}
- \hat{y}_e z_{\mu}^{(3)} \hat{\beta}_\mu - z_e^{(2)} \hat{y}_\mu \hat{\beta}_\mu) \, t^2 \, u^2 ~~,\nonumber\\
\mathcal{\hat M}_{e\tau}&= \frac{1}{\hat{y}_{\tau}^2} \, (z_\tau^{(2)} \hat{y}_{\tau} \hat{\beta}_e + \hat{y}_{\tau}^{2} \gamma_e ^{(3)}
-\hat{y}_{e} z_\tau^{(2)} \hat{\beta}_\tau - z_e ^{(3)} \hat{y}_\tau \hat{\beta}_\tau) \, t^2 \, u^2~~,\nonumber\\
\mathcal{\hat M}_{\mu e}&= \frac{1}{\hat{y}_\mu} (\hat{y}_\mu \gamma_\mu ^{(3)} - z_\mu ^{(3)} \hat{\beta}_\mu) \, t\, u^2 ~~,\nonumber\\
\label{Mhat}\mathcal{\hat M}_{\mu\mu}&= \hat{\beta}_\mu \, t \, u ~~,\\
\mathcal{\hat M}_{\mu\tau}&= \frac{1}{\hat{y}_\tau^2} \, (\hat{y}_\tau z_\tau ^{(3)} \hat{\beta}_\mu + \hat{y}_\tau^2 \gamma_\mu ^{(2)}
-z_\tau ^{(3)} \hat{y}_\mu \hat{\beta}_\tau - z_\mu ^{(2)} \hat{y}_\tau \hat{\beta}_\tau) \, t \, u^2~~,\nonumber\\
\mathcal{\hat M}_{\tau e}&= \frac{1}{\hat{y}_\tau} \, (\hat{y}_\tau \gamma_\tau ^{(2)} - z_\tau ^{(2)} \hat{\beta}_\tau) \, u^2~~,\nonumber
\end{align}
\begin{align}
\mathcal{\hat M}_{\tau\mu}&= \frac{1}{\hat{y}_\tau} \, (\hat{y}_\tau \gamma_\tau ^{(3)} - z_\tau ^{(3)} \hat{\beta}_\tau) \, u^2 ~~,\nonumber\\
\mathcal{\hat M}_{\tau\tau}&= \hat{\beta}_\tau \, u ~~.\nonumber
\end{align}
The results shown here have been obtained under the assumption that all parameters are real, as explained in appendix B,
but they remain unchanged in the complex case as far as the diagonal elements of $\hat{\cal M}$ are concerned.
Notice that, in general, all entries of $\mathcal{\hat M}$, after applying the appropriate
transformations for canonical normalization of the kinetic terms and diagonalization
of the charged lepton mass matrix, are of the same order as the corresponding entries of $\mathcal{M}$ before
the transformations, see eq. (\ref{M_subleading}). This is expected, if no additional
cancellation or enhancement takes place. Note further that
the structure of the leading order result is surprisingly simple, since it does not involve parameters
appearing in the non-canonically normalized kinetic terms. By having a closer look at the explicit form of ${\cal \hat{M}}_{ij}$ one sees that the elements
above the diagonal of $\hat{\cal M}$, $\hat{\cal M}_{e\mu}$, $\hat{\cal M}_{e\tau}$ and $\hat{\cal M}_{\mu\tau}$,
have a similar structure. Similarly, the elements $\hat{\cal M}_{ij}$
below the diagonal, $\hat{\cal M}_{\mu e}$, $\hat{\cal M}_{\tau e}$ and $\hat{\cal M}_{\tau\mu}$, have a common structure. 
Finally, note that for $c_2=c_3$,
 which implies $z_{f}^{(2)}=z_{f}^{(3)}$ and $\gamma_{f}^{(2)}=\gamma_{f}^{(3)}$, the elements
 ${\cal \hat{M}}_{\tau e}$ and ${\cal \hat{M}}_{\tau \mu}$ become equal. 
The MDMs and EDMs are given at lowest order as:
\bea
a_e &= 2 m_e\dd\frac{v}{\sqrt{2} M^2} Re(\hat{\beta}_{e} \, t^2 \, u)~~,\nonumber\\
\label{MDM}
a_\mu &= 2 m_\mu\dd\frac{v}{\sqrt{2} M^2} Re(\hat{\beta}_{\mu} \, t \, u)~~,\\
a_\tau &= 2 m_\tau\dd\frac{v}{\sqrt{2} M^2} Re(\hat{\beta}_{\tau} \, u)~~,\nonumber
\eea
and
\bea
d_e&=e \dd\frac{v}{\sqrt{2} M^2} Im (\hat{\beta}_e  t^2 u)~~,\nonumber\\
\label{EDM}
d_\mu&=e \dd\frac{v}{\sqrt{2} M^2} Im (\hat{\beta}_\mu  t u)~~,\\
d_\tau&=e \dd\frac{v}{\sqrt{2} M^2} Im (\hat{\beta}_\tau u)~~.\nonumber
\eea
As we have anticipated, MDMs and EDMs arise at first order in the $u$ parameter. As a consequence their expressions
in eqs. (\ref{MDM}) and (\ref{EDM}) depend neither on the parameters ${\hat y}_f$ and $z^{(k)}_f$ of the charged lepton mass matrix, nor on the off-diagonal
elements of ${\cal M}$, but only on ${\hat\beta}_f$, at the leading order.
Notice that, for coefficients $\hat{\beta}_f$ with absolute values and phases of order one, as expected in our
model, we have:
\be
a_i=O\left(2\dd\frac{m_i^2}{M^2}\right)~~~,~~~~~~~~~~d_i=O\left(e\dd\frac{m_i}{M^2}\right)
\label{oom}
\ee
We can derive a bound on the scale $M$, by considering the
existing limits on MDMs and EDMs and by using eqs. (\ref{oom}) as exact equalities to fix the ambiguity of the unknown coefficients $\hat\beta_f$.
We find the results shown in table 2.
\begin{table}[!ht]
\centering
                \begin{math}
                \begin{array}{|c|c|}
                    \hline
                    & \\[-9pt]
                    d_e<1.6\times 10^{-27}~~e~cm&M>80~~{\rm TeV}\\[3pt]
                    \hline
                    &\\[-9pt]
                   d_\mu<2.8\times 10^{-19}~~e~cm&M>80~~{\rm GeV}\\[3pt]
                    \hline
                    &\\[-9pt]
                    \delta a_e<3.8\times 10^{-12}&M>350~~{\rm GeV}\\[3pt]
                    \hline
                    &\\[-9pt]
                    \delta a_\mu\approx 30\times 10^{-10}&M\approx 2.7~~{\rm TeV}\\[3pt]
                    \hline
                \end{array}
               \end{math}
            \caption{Experimental limits on lepton MDMs and EDMs\cite{ExperimentalBoundsMDMmu,ExperimentalBoundsMDMe,PDG} and corresponding bounds on the scale $M$, derived from eqs. (\ref{oom}).
The data on the $\tau$ lepton have not been reported since they are much less constraining. For the anomalous magnetic moment of the muon,
$\delta a_\mu$ stands for the deviation of the experimental central value from the SM expectation.}
            \end{table}
\vskip 0.2cm
\noindent
We see from table 2 that, in order to accept values of $M$ in the range $1\div 10$ TeV, we should invoke a cancellation in the imaginary part
of $\hat{\cal{M}}_{ee}$, which can be either accidental or due to CP-conservation in the considered sector of the theory.
Concerning the flavour violating dipole transitions, using eq. ({\ref{Mhat}) we see that the rate for $l_i\to l_j\gamma$ is dominated by
the contribution ${\cal\hat M}_{ij}$, since ${\cal\hat M}_{ji}$ is suppressed by a relative factor of $O(t)$ for $\mu\to e\gamma $ and $\tau\to\mu\gamma$
and of $O(t^2)$ for $\tau\to e\gamma$. We get:
\be
\frac{BR(l_i\to l_j\gamma)}{BR(l_i\to l_j\nu_i{\bar \nu_j})}=\frac{48\pi^3 \alpha}{G_F^2 M^4}\vert w_{ij} ~u\vert^2
\label{LFV}
\ee
where
\bea
w_{\mu e}&=&\dd\frac{1}{{\hat y}_\mu^2}\left({\hat y}_\mu\gamma_\mu^{(3)}-z_\mu^{(3)} {\hat\beta}_\mu\right)\nonumber\\
w_{\tau \mu}&=&\dd\frac{1}{{\hat y}_\tau^2}\left({\hat y}_\tau\gamma_\tau^{(3)}-z_\tau^{(3)} {\hat\beta}_\tau\right)\nonumber\\
w_{\tau e}&=&\dd\frac{1}{{\hat y}_\tau^2}\left({\hat y}_\tau\gamma_\tau^{(2)}-z_\tau^{(2)} {\hat\beta}_\tau\right)~~~.
\label{wij}
\eea
The form of $w_{ij}$ shows that for each transition there are two contributions to the rate: one, proportional to $\gamma^{(2,3)}_i$, coming from the off-diagonal element $(ij)$
of the original dipole matrix ${\cal M}$ and one, proportional to $z^{(2,3)}_i$, coming from the effect of diagonalizing the charged lepton mass matrix,
see eqs. (\ref{M_subleading}) and (\ref{yl_subleading}), respectively.
Both contribute at the same order in $u$.
Since the quantities $w_{ij}$ are all expected to be of order one, we can conclude that in the class of models considered here the branching ratios for the three
transitions $\mu\to e\gamma $, $\tau\to\mu\gamma$ and $\tau\to e\gamma$ should all be of the same order:
\be
BR(\mu\to e \gamma)\approx BR(\tau\to\mu\gamma)\approx BR(\tau\to e \gamma)~~~.
\ee
This is a distinctive feature of our class of models, since in most of the
other existing models there is a substantial difference between the branching ratios\cite{MFV,MLFV,SUSYLFV+symmetries,SUSYFP+GUTs}. In particular it often occurs that
$BR(\mu\to e \gamma)<BR(\tau\to\mu\gamma)$. Given the present experimental bound $BR(\mu\to e \gamma)<1.2\times 10^{-11}$\cite{ExperimentalBoundsDecaysMu}, our result implies that $\tau\to\mu\gamma$ and $\tau\to e \gamma$ have rates much below the present and expected future sensitivity\cite{ExperimentalBoundsDecaysTau}.
Moreover, from the current (future) experimental limit on $BR(\mu\to e \gamma)$\cite{ExperimentalBoundsDecaysMu} and assuming $\vert w_{\mu e}\vert=1$, we derive the following bound
on $\vert u/M^2\vert$:
\be
BR(\mu\to e \gamma)<1.2\times 10^{-11}~(10^{-13})~~~~~~~
\left\vert\dd\frac{u}{M^2}\right\vert<1.2\times 10^{-11}~(1.1\times 10^{-12})~~{\rm GeV}^{-2}~~~.
\ee
\noindent
Since the parameter $\vert u\vert$ lies in the limited range $0.001<\vert u \vert<0.05$, see eq. (\ref{ubound}), we find
\bea
\vert u\vert =& 0.001 ~~~~~~~~~~~~~~&M>10~(30)~~{\rm TeV}\\
\vert u\vert =& 0.05 ~~~~~~~~~~~~~~&M>70~(200)~~{\rm TeV}~~~.
\eea
This pushes the scale $M$ considerably above the range we were initially interested in. In particular $M$ is shifted above the region of
interest for $(g-2)_\mu$ and probably for LHC.

Since $\hat{{\cal M}}_{\tau e}$ and $\hat{{\cal M}}_{\tau\mu}$ become equal for $c_2=c_3$, as
explained above, also $w_{\tau e}$ and $w_{\tau\mu}$ in eq. (\ref{wij}) become equal and
therefore also the branching ratio of $\tau \to e \gamma$ and $\tau \to \mu \gamma$
will be the same, at the leading order. 
%
%
\section{Supersymmetric case}
As we have seen, the off-diagonal elements of the dipole matrix ${\cal M}$ can be traced back to two independent sources.
They can originate either from ${\cal L}_{dip}^{(1)}$, when the sub-leading corrections to the VEV of the $\varphi_T$
multiplet (terms proportional to $c_{1,2,3}$ in eq. (\ref{vevs})) are accounted for or from ${\cal L}_{dip}^{(2)}$, 
where the relevant double flavon insertions (see eq. (\ref{Xlist})) are considered.
In this second case, the only combinations of flavon insertions that can provide a non-vanishing contribution are
$\xi^\dagger \varphi_S$ and its conjugate. In a generic case we expect that both these contributions are equally important
and contribute at the same order to a given off-diagonal dipole transition.
There is however a special case where the double flavon insertions $\xi^\dagger \varphi_S$ and its conjugate
are suppressed compared to the sub-leading corrections to $\varphi_T$ and an overall depletion in the
elements of the matrix $\hat{\cal M}$ below the diagonal takes place. This happens, under certain conditions that
we are going to specify, when the underlying theory is supersymmetric and supersymmetry is softly broken.

In order to understand the reason why in the supersymmetric case the insertions $\xi^\dagger \varphi_S$ and its conjugate
are suppressed, it is useful to recall what are the expected sources of chirality flip in the lepton sector.
In the supersymmetric case we have two Higgs doublets, $H_{u,d}$, and
the field $H^\dagger$ of section 2 is replaced by $H_d$. Moreover, in the limit of unbroken supersymmetry
the lepton mass terms are derived from the superpotential, which is holomorphic in the chiral superfields.
At the leading order in the flavour symmetry breaking parameters, these terms are of the type:
\be
\frac{1}{\Lambda_f}\int d^2\theta_{SUSY}  \tau^c H_d (\varphi_T l)''~~~
\ee
and similarly for the other charged fermions.
Therefore the terms in ${\cal L}^{(1,2)}_l$ containing non-holomorphic fields like $\varphi_T^\dagger$
and $X_i$ $(i=3,...,7)$ vanish in the supersymmetric limit:
\be
y'_f=0~~~,~~~~~~~y^{k}_f=0~~~(k=3,...,7)
\label{exsusy}
\ee
As a matter of fact, in a realistic model supersymmetry is broken,
for instance by the VEV $F$ of the $\theta^2_{SUSY}$ component of a chiral superfield.
Here we assume that supersymmetry breaking originates in a hidden sector of the theory and is transmitted to the observable sector
via interactions suppressed by the scale $\Lambda_f$ or by a larger scale,
giving rise to soft supersymmetry breaking terms with a typical mass scale of order $m_{SUSY}$.
Supersymmetry breaking can give rise to non-holomorphic contributions to lepton masses,
which are however suppressed by powers of $m_{SUSY}/\Lambda_f$, as in the following example:
\be
\frac{1}{\Lambda_f^2}\int d^2\theta_{SUSY} d^2 \overline{\theta}_{SUSY} \tau^c H_d (\varphi_T^\dagger l)''~ \overline{\theta}^2_{SUSY} m_{SUSY}
\ee
The ratio $m_{SUSY}/\Lambda_f$ is a tiny quantity, which we will neglect in our analysis.

Beyond fermion mass terms another source of chirality flip is provided by the slepton
mass terms of left-right type, which involve both an SU(2) doublet and an SU(2) singlet slepton.
At the leading order in the supersymmetry breaking parameter $m_{SUSY}$, such soft breaking masses
do not contain any insertion of anti-holomorphic fields like $\varphi_{T,S}^\dagger$ and $\xi^\dagger$.
Indeed insertions of holomorphic supermultiplets only require one power of $m_{SUSY}$ as in the following example
\be
\frac{1}{\Lambda_f}\int d^2\theta_{SUSY}  \tau^c H_d (\varphi_T l)''~ \theta^2_{SUSY} m_{SUSY}~~~.
\ee
On the contrary the insertion of an anti-holomorphic supermultiplet requires two powers of $m_{SUSY}$:
\be
\frac{1}{\Lambda_f^2}\int d^2\theta_{SUSY}  d^2 \overline{\theta}_{SUSY} \tau^c H_d (\varphi_T^\dagger l)''~ \theta^2_{SUSY} \overline{\theta}^2_{SUSY} m_{SUSY}^2~~~.
\label{antih}
\ee
In our approximation we will also neglect all contributions of the type in eq. (\ref{antih}). Furthermore,
we assume that in the underlying fundamental theory the only sources of chirality flip are
either fermion masses or sfermion masses of left-right type. Both of them, up to the order $u^2$,
are described by the insertion of $\varphi_T$ or $\varphi_T^2$ in the relevant operators.
This is our definition of the ''supersymmetric case''.
Therefore in the supersymmetric case we should 
use the restrictions shown in eqs. (\ref{exsusy}).
The fact that any chirality flip up to the order $u^2$ necessarily requires the insertion of $\varphi_T$ or
$\varphi_T^2$ also applies to the operators of ${\cal L}_{dip}$, and we have
(analogously to eq. (\ref{exsusy})):
\be
\beta'_f=0~~~,~~~~~~~\beta^{k}_f=0~~~(k=3,...,7)~~~.
\label{betaexsusy}
\ee
By applying eqs. (\ref{exsusy}) and (\ref{betaexsusy}) we find the following matrices $y_l$ and $\mathcal{M}$ in the supersymmetric case:
\begin{equation}
\label{yl_subleading_SUSY}
y_{l} = \left( \begin{array}{ccc}
        y_{e} t^2 u + (y_e \, c_1 + y_e^2) \, t^2 u^2 & y_e \, c_3 \, t^2 u^2 &  y_e \, c_2 \, t^2 u^2\\
        y_\mu \, c_2 \, t u^2 & y_{\mu} t u + (y_\mu \, c_1 + y_\mu^2) \, t u^2 & y_\mu \, c_3 \, t u^2\\
        y_\tau \, c_3 \, u^2 & y_\tau \, c_2 \, u^2 & y_{\tau} u + (y_\tau \, c_1 + y_\tau ^{2}) \, u^2
\end{array}
\right)
\end{equation}
and
\begin{equation}
\label{M_subleading_SUSY}
\mathcal{M} = \left( \begin{array}{ccc}
        \beta_{e} \, t^2 u + (\beta_e \, c_1 + \beta_e ^2) \, t^2 u^2 & \beta_{e}\, c_3 \, t^2 u^2 &  \beta_{e}\, c_2 \, t^2 u^2\\
        \beta_{\mu} \, c_2 \, t u^2 & \beta_\mu \, t u + (\beta_{\mu} \, c_1 + \beta_\mu^2) \, t u^2 &  \beta_{\mu} \, c_3 \, t u^2\\
        \beta_{\tau} \, c_3 \, u^2 & \beta_{\tau} \, c_2 \, u^2 & \beta_{\tau} \, u + (\beta_\tau \, c_1 + \beta_\tau ^2) \, u^2
\end{array}
\right)
\end{equation}
The procedure to arrive at $\mathcal{\hat M}$ is exactly the same as before. Thereby, note that the
form of the kinetic terms $K$ and $K^c$ is the same as before, since supersymmetry puts no restriction
on these terms.
The result for the entries of $\mathcal{\hat M}$ is then\footnote{Note that if we take the formulae given in eqs. (\ref{Mhat}) and use eqs. (\ref{exsusy})
and (\ref{betaexsusy}) we also arrive at this form of ${\cal \hat{M}}_{ij}$. Thereby, the elements
${\cal \hat{M}}_{\mu e}$, ${\cal \hat{M}}_{\tau e}$ and ${\cal \hat{M}}_{\tau\mu}$
below the diagonal turn out to be zero, i.e. we can also deduce the suppression of
these elements, however not their exact form.}:
\begin{align}
\mathcal{\hat M}_{ee}&= \beta_e \, t^2 \, u~~,\nn\\
\mathcal{\hat M}_{e\mu}&=  \frac{c_2 +c_3}{y_\mu} \, (\beta_e \, y_\mu - y_e \, \beta_\mu) \, t^2 \, u^2 ~~,\nn\\
\mathcal{\hat M}_{e\tau}&= \frac{c_2+c_3}{y_\tau} \, (\beta_e \, y_\tau - y_e \, \beta_\tau) \, t^2 \, u^2 ~~,\nn\\
\mathcal{\hat M}_{\mu e}&=  \frac{c_2}{y_\mu} \, (y_\mu ^2 \, \beta_\mu - \beta_\mu^2 \, y_\mu) \, t \, u^3~~,\nn\\
\mathcal{\hat M}_{\mu\mu}&= \beta_\mu \, t \, u ~~,\\
\mathcal{\hat M}_{\mu\tau}&= \frac{c_2+c_3}{y_\tau} \, (\beta_\mu \, y_\tau - \beta_\tau \, y_\mu) \, t \, u^2 ~~,\nn\\
\mathcal{\hat M}_{\tau e}&= \frac{c_3}{y_\tau} \, (y^2_\tau \, \beta_\tau - \beta_\tau^2 \, y_\tau) \, u^3 ~~,\nn\\
\mathcal{\hat M}_{\tau\mu}&=  \frac{c_2}{y_\tau} \, (y_\tau ^2 \, \beta_\tau - \beta_\tau^2 \, y_\tau) \, u^3 ~~,\nn\\
\mathcal{\hat M}_{\tau\tau}&= \beta_\tau \, u~~.\nn
\end{align}
Again, the leading order results are not affected by the non-canonically normalized kinetic terms
and therefore are rather simple.
Also here, the elements
above the diagonal $\hat{\cal M}_{ij}$ have a common structure, i.e. they are determined by the
sum of $c_2$ and $c_3$ and the parameters $y_f$ and $\beta_{f^{\prime}}$ where indices $f$ and $f^{\prime}$
are determined by $i$ as well as $j$. Similarly, for the elements ${\cal \hat{M}}_{ij}$ below the
diagonal we find that they are either proportional to $c_2$ or $c_3$ and the index $f$ of all parameters is solely given by the index $i$ of ${\cal \hat{M}}_{ij}$. In this case the
equality of ${\cal \hat{M}}_{\tau e}$ and ${\cal \hat{M}}_{\tau\mu}$ for $c_2=c_3$ also holds and
is more obvious than in the general case.
The EDMs and MDMs are similar to those of the general non-supersymmetric case:
we find the same degree of suppression in the $t$ and $u$ parameters for the physical quantities.
The difference between the general and the supersymmetric approaches becomes manifest  when the sub-leading
corrections are taken into account, i.e. only in the study of the LFV processes.
The main difference is the suppression of the elements below the diagonal which are not of the order as expected from eq. (\ref{M_subleading_SUSY}), but
suppressed by an additional power of $u$. We get
\be
\frac{BR(l_i\to l_j\gamma)}{BR(l_i\to l_j\nu_i{\bar \nu_j})}= \frac{48\pi^3 \alpha}{G_F^2 M^4}\left[\vert w^{(1)}_{ij} u^2\vert^2+\frac{m_j^2}{m_i^2} \vert w^{(2)}_{ij} u\vert^2\right]
\label{LFVsusy}
\ee
where
\bea
w^{(1)}_{\mu e}&=\dd\frac{c_2}{y_\mu^2} \, (y_\mu ^2 \, \beta_\mu - \beta_\mu^2 \, y_\mu)~~~~~~~~~~&
w^{(2)}_{\mu e}=\frac{c_2 +c_3}{y_\mu y_e} \, (\beta_e \, y_\mu - y_e \, \beta_\mu)\\
w^{(1)}_{\tau\mu}&=\dd\frac{c_2}{y_\tau^2} \, (y_\tau ^2 \, \beta_\tau - \beta_\tau^2 \, y_\tau)~~~~~~~~~~&
w^{(2)}_{\tau\mu}=\frac{c_2+c_3}{y_\tau y_\mu} \, (\beta_\mu \, y_\tau - \beta_\tau \, y_\mu)\\
w^{(1)}_{\tau e}&=\dd\frac{c_3}{y_\tau^2} \, (y^2_\tau \, \beta_\tau - \beta_\tau^2 \, y_\tau)~~~~~~~~~~&
w^{(2)}_{\tau e}=\frac{c_2+c_3}{y_\tau y_e} \, (\beta_e \, y_\tau - y_e \, \beta_\tau)
\label{wsusy}
\eea
\begin{figure}[t]
\begin{center}
        \mbox{\epsfig{figure=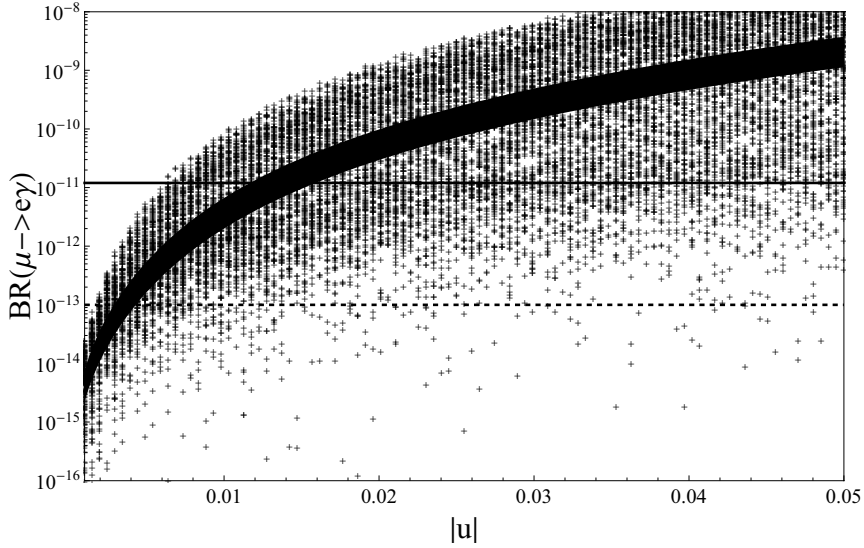,width=12.0cm}}
\end{center}
\caption{The branching ratio of $\mu\to e \gamma$ as a function of $|u|$, eq. (75). The deviation of the anomalous magnetic moment of the muon from the SM expectation is kept fixed to its experimental range\cite{ExperimentalBoundsMDMmu}.
The unknown coefficients $\tilde{w}^{(1,2)}_{\mu e}$ are equal to 1 (darker region) or are random
complex numbers with absolute values between zero and two (lighter region).
The continuous (dashed) horizontal line corresponds to the present (future expected) experimental bound on $BR(\mu\to e\gamma)$\cite{ExperimentalBoundsDecaysMu}.}
\label{BR}
\end{figure}
Notice that now the first contribution on the right-hand side of eq. (\ref{LFVsusy}) is suppressed by a factor of $u$ compared to the non-supersymmetric case.
In most of the allowed range of $u$, the branching ratios of $\mu\to e \gamma$ and $\tau \to\mu \gamma$ are similar and larger than
the branching ratio of $\tau\to e \gamma$.
Assuming $\vert w^{(1,2)}_{\mu e}\vert=1$,
the present (future) experimental limit on $BR(\mu\to e \gamma)$ implies the following bounds
\begin{eqnarray}
\vert u\vert =& 0.001 ~~~~~~~~~~&M>0.7~(2)~~{\rm TeV}\\
\vert u\vert =& 0.05 ~~~~~~~~~~~~~~&M>14~(48)~~{\rm TeV}~~~.
\end{eqnarray}
We see that at variance with the non-supersymmetric case there is a range of permitted values of the parameter $|u|$
for which the scale $M$ can be sufficiently small to allow an explanation of the observed discrepancy in $a_\mu$,
without conflicting with the present bound on $BR(\mu\to e \gamma)$.
We can eliminate the dependence on the unknown scale $M$ by combining eqs. (\ref{oom}) and (\ref{LFVsusy}). For $\mu\to e \gamma$
we get\footnote{Relations among $BR(\mu\to e \gamma)$, $\delta a_\mu$ and neutrino mass parameters have also been discussed in
\cite{g2BR}.}:
\begin{equation}
\frac{BR(\mu\to e\gamma)}{BR(\mu\to e\nu_\mu{\bar \nu_e})}= \frac{12\pi^3 \alpha}{G_F^2 m_\mu^4}\left(\delta a_\mu\right)^2
\left[\vert \tilde{w}^{(1)}_{\mu e}\vert^2 \vert u\vert^4+\frac{m_e^2}{m_\mu^2} \vert \tilde{w}^{(2)}_{\mu e}\vert^2\vert u\vert^2\right]
\label{muegamma}
\end{equation}
where $\tilde{w}^{(1,2)}_{\mu e}$ are unknown, order one coefficients. We plot $BR(\mu\to e\gamma)$ versus $|u|$ in fig. 1,
where the coefficients $\tilde{w}^{(1,2)}_{\mu e}$ are kept fixed to 1 (darker region) or are random
complex numbers with absolute values between zero and two (lighter region). The deviation of the anomalous magnetic moment
of the muon from the SM prediction is in the interval of the experimentally allowed values,
about three sigma away from zero. Even if the ignorance about the coefficients $\tilde{w}^{(1,2)}_{\mu e}$ does not allow us
to derive a sharp limit on $|u|$, we see that the present limit on $BR(\mu\to e \gamma)$ disfavors values of $|u|$
larger than few percents. We recall that in this model the magnitudes of $|u|$ and $\theta_{13}$ are comparable.
%
\section{Comparison with Minimal Flavour Violation}
It is instructive to compare the previous results with those of MFV \cite{MFV,MLFV}.
If we restrict ourselves to the case where right-handed neutrinos do not affect
the flavour properties of the theory, the flavour group of MFV is $G_f=SU(3)_{e^c}\times SU(3)_l\times ...$,
where we have only displayed the part relevant for the lepton sector. Electroweak singlets $e^c$ and doublets $l$ transform as
$(3,1)$ and $(1,\bar{3})$, respectively. The flavon fields or, better, their VEVs are the Yukawa couplings $y_l$ and $Y$ themselves,
transforming as $(\bar{3},3)$ and $(1,6)$, respectively. By going to a basis where the charged leptons are diagonal, $y_l$ and $Y$
can be expressed in terms of lepton masses and mixing angles (we keep using a notation where a hat denotes matrices in this particular
basis):
\begin{equation}
\hat{y}_l=\frac{\sqrt{2}}{v} m_l^{\rm diag}~~~,~~~~~~~~~~\hat{Y}=\frac{\Lambda_L}{v^2} U^* m_\nu^{\rm diag} U^\dagger~~~,
\end{equation}
where $U$ is the lepton mixing matrix.
The diagonal elements of the matrix $\hat{\cal M}$ evaluated in MFV are analogous to those of the previous class of models and similar bounds
on the scale $M$ are derived from the existing data on MDMs and EDMs. The off-diagonal elements are given by:
\begin{eqnarray}
\hat{\cal M}_{ij}&=&\beta (\hat{y}_l \hat{Y}^\dagger \hat{Y})_{ij}\nonumber\\
&=&\sqrt{2}\beta\frac{m_i}{v}\frac{\Lambda_L^2}{v^4}\left[\Delta m^2_{sol} U_{i2} U^*_{j2}\pm\Delta m^2_{atm} U_{i3} U^*_{j3}\right]
\end{eqnarray}
where $\beta$ is an overall coefficient of order one and the plus (minus) sign refers to the case of normal (inverted) hierarchy.
We see that, due to the presence of the ratio $\Lambda_L^2/v^2$ the overall scale of these matrix elements is much less constrained than in the previous case.
This is due to the fact that MFV does not restrict the overall strength of the coupling constants $Y$, apart from the requirement
that they remain in the perturbative regime.
Very small or relatively large (but smaller than one) $Y$ can be accommodated by adjusting the scale $\Lambda_L$.
On the contrary this is not allowed in the case previously discussed where the size of the symmetry breaking effects is restricted to the small window (\ref{ubound})
and the scale $\Lambda_L$ is determined within a factor of about fifty.
The conclusion is that in MFV the non-observation of $l_i\to l_j \gamma$ could be justified by choosing a small $\Lambda_L$, while a positive signal
in $\mu\to e \gamma$ with a branching ratio in the range $1.2\times 10^{-11}\div 10^{-13}$ could also be fitted by an appropriate $\Lambda_L$,
apart from a small region of the $\theta_{13}$ angle, around $\theta_{13}\approx0.02$ where a cancellation can take place.

The dependence on the scale $\Lambda_L$ is eliminated by considering ratios of branching ratios. For instance:
\begin{equation}
\frac{BR(\mu\to e\gamma)}{BR(\mu\to e\nu_\mu{\bar \nu_e})}\frac{BR(\tau\to \mu\nu_\tau{\bar \nu_\mu})}{BR(\tau\to \mu\gamma)}=
\left\vert\frac{2\Delta m^2_{sol}}{3\Delta m^2_{atm}}\pm \sqrt{2}\sin\theta_{13} e^{i\delta}\right\vert^2<1~~~,
\label{mfv}
\end{equation}
where we took the TB ansatz, see eq. (\ref{thetatb}), to fix $\theta_{12}$ and $\theta_{23}$.
We see that $BR(\mu\to e\gamma)<BR(\tau\to \mu\gamma)$ always in MFV. Moreover, for $\theta_{13}$ above approximately $0.07$, $BR(\mu\to e\gamma)<1.2\times 10^{-11}$ implies $BR(\tau\to \mu\gamma)<10^{-9}$. For $\theta_{13}$ below $0.07$, apart possibly from a small region around $\theta_{13}\approx0.02$, both the transitions $\mu\to e \gamma$ and $\tau\to\mu\gamma$
might be above the sensitivity of the future experiments.

We also observe that in MFV the only difference between the general case and the supersymmetric one is the presence of two doublets
in the low-energy Lagrangian. In MFV a chirality flip in leptonic operators necessarily requires the insertion of the matrix $\hat{y}_l$,
both in the general and in the supersymmetric case and, apart from the possibility of $\tan\beta$ enhanced contributions, 
similar predictions for the LFV processes are expected in the two cases. 
%
\section{Conclusion}
Flavour mixing within lepton families is surprisingly close to the TB one\cite{TB}, with a striking agreement for the solar and the atmospheric mixing angles.
The closeness of the lepton mixing to the TB one is still waiting for a full experimental confirmation, which could be provided by the evidence for a relatively small
angle $\theta_{13}$, of the order of few degrees. In the recent years several models of lepton masses and mixing angles
predicting an approximate TB mixing have been proposed\cite{TBA4,af1,af2,afl,afh,linyin,TBother}. A particularly simple model is the one based on the flavour symmetry $A_4\times Z_3\times U(1)_{FN}$\cite{af1,af2,afl},
which represents a sort of minimal choice to properly account for both, the neutrino and the charged lepton sector.
The TB mixing originates from the misalignment of the scalar multiplets $\varphi$ that break $A_4$ giving masses to neutrinos and to charged leptons.
In this paper we have analyzed a set of new low-energy predictions of models
invariant under the flavour symmetry $A_4\times Z_3\times U(1)_{FN}$.  These predictions involve leptonic MDMs, EDMs
and LFV transitions like $\mu\to e \gamma$, $\tau\to \mu\gamma$ and $\tau \to\ e \gamma$.
We have constructed the effective low-energy Lagrangian that describes these observables.
Such an effective Lagrangian is invariant under the flavour symmetry, and all flavour breaking effects are encoded
in the dependence on the spurions $\Phi\equiv\varphi/\Lambda_f$ ($\Lambda_f$ being the cut-off of the theory).
These also control lepton masses and mixing angles.
The dominant operators are obtained by expanding the Lagrangian in powers of $\Phi$ and by keeping the first few terms
in the expansion. The leading contributions have dimension six and are suppressed by two powers of a new scale $M$.
This scale can be considerably smaller than the other scales of the problem, like the cut-off $\Lambda_f$,
the VEV of the flavons $\langle\varphi\rangle$ and the scale of lepton number violation $\Lambda_L$. We treat $M$ as an independent parameter,
to be constrained by the experiments. Apart from the scale $M$, all the relevant information needed to
predict MDMs, EDMs and LFV transitions is contained in a dimensionless
matrix $\hat{\cal M}$, whose elements can be computed up to unknown order-one coefficients from our Lagrangian.
The strongest bound comes from the EDM of the electron\cite{ExperimentalBoundsMDMe}
: $M>80$ TeV.
A lower value for $M$ can be tolerated in the presence of a cancellation in the imaginary part
of $\hat{\cal{M}}_{ee}$, perhaps related to CP-conservation in the considered sector of the theory.
This problem is also present in MFV\cite{MFV,MLFV}, where one simply assumes that $\hat{\cal{M}}_{ee}$ is real.
Another stringent bound can be derived from the analysis of the $\tau$ decays $\tau^-\to \mu^+ e^- e^-$ and $\tau^-\to e^+ \mu^- \mu^-$\cite{PDG}
that, barring cancellations in the coefficients of the relevant operators, push the scale $M$ above approximately 15 TeV.
Coming to LFV dipole transitions, we have found that in the general case the branching ratios for
$\mu\to e \gamma$, $\tau\to \mu\gamma$ and $\tau \to\ e \gamma$ are all expected to be of the same order, at variance with MFV.
Given the present limit on $BR(\mu\to e \gamma)$\cite{ExperimentalBoundsDecaysMu}, this implies that $\tau\to \mu\gamma$ and $\tau \to\ e \gamma$
have rates much smaller than the present (and near future) sensitivity\cite{ExperimentalBoundsDecaysTau}.
The absolute values of these branching ratios depend on the spurion VEVs $\langle\Phi\rangle$ that in our class of models
are determined by a parameter $|u|$ in the range $0.001<|u|<0.05$. In the general case, for $BR(\mu\to e \gamma)<1.2\times 10^{-11}~(10^{-13})$
we get $M>10~(30)$ TeV if $|u|=0.001$ and $M>70~(200)$ TeV for $|u|=0.05$.
The anomalous MDM of the muon $a_\mu$ and its deviation from the SM expectation\cite{ExperimentalBoundsMDMmu} provide the indication for a lower scale $M$,
of the order of few TeV, which would also be of great interest for LHC.
In order to reconcile this possibility with the results derived from the LFV dipole transitions, we have
reconsidered the matrix $\hat{\cal M}$ in a supersymmetric context, where additional
constraints have to be applied. The operators describing $\mu\to e \gamma$, $\tau\to \mu\gamma$ and $\tau \to\ e \gamma$
flip the lepton chirality. By assuming that in a supersymmetric theory the only sources of chirality flips are the fermion masses and the
sfermion mass terms of left-right type, we find that a cancellation takes place in the elements of $\hat{\cal M}$
below the diagonal. As a result the limits on the scale $M$ become less severe.
For $BR(\mu\to e \gamma)<1.2\times 10^{-11}~(10^{-13})$
we get $M>0.7~(2)$ TeV if $|u|=0.001$ and $M>14~(48)$ TeV for $|u|=0.05$.
At variance with the non-supersymmetric case there is a range of values of the parameter $|u|$
for which the scale $M$ can be sufficiently small to allow for an explanation of the observed discrepancy in $a_\mu$,
without conflicting with the present bound on $\mu\to e \gamma$. Since in our framework $\theta_{13}$ is comparable to $|u|$,
the present limit on $BR(\mu\to e \gamma)$ together with the existing discrepancy
in $a_\mu$ point to a rather small value for $\theta_{13}$, of the order of few percents in radians,
close to but probably just below the sensitivity expected in future experiments at reactors or with high intensity neutrino beams\cite{Theta13future,Schwetz:2007my}.
It remains to be seen if the bounds on the scale $M$ derived from four-lepton operators can be evaded in the supersymmetric case.
\section*{Acknowledgments}
We acknowledge useful discussions with Paolo Gambino, Michael A. Schmidt and with Fabio Zwirner.
C.H. would like to thank the theory group of the University of
Padua and the INFN section of Padua for very kind hospitality.
We recognize that this work has been partly supported by the European Commission under contracts MRTN-CT-2004-503369 and MRTN-CT-2006-035505
and by the ``Sonderforschungsbereich'' TR27.
\vfill

\vfill
\newpage
%
%
\section*{A~~~The group $A_4$}
The group $A_4$ is generated by two elements $S$ and $T$ obeying the relations\cite{A4Presentation}:
\be
S^2=(ST)^3=T^3=1~~~.
\ee
It has three independent one-dimensional representations, $1$, $1'$ and $1''$ and one three-dimensional representation $3$.
The one-dimensional representations are given by:
\be
\begin{array}{lll}
1&S=1&T=1\\
1'&S=1&T=e^{\dd i 4 \pi/3}\equiv\omega^2\\
1''&S=1&T=e^{\dd i 2\pi/3}\equiv\omega\\
\label{s$A_4$}
\end{array}
\ee
The three-dimensional representation, in a basis
where the generator $T$ is diagonal, is given by:
\be
T=\left(
\begin{array}{ccc}
1&0&0\\
0&\omega^2&0\\
0&0&\omega
\end{array}
\right),~~~~~~~~~~~~~~~~
S=\frac{1}{3}
\left(
\begin{array}{ccc}
-1&2&2\cr
2&-1&2\cr
2&2&-1
\end{array}
\right)~~~.
\label{ST}
\ee

The multiplication rule for triplet representations is the following:
\be
3\times 3=1+1'+1''+3_S+3_A
\ee
If we denote by
\be
a=(a_1,a_2,a_3)~~~,~~~~~~~~~~b=(b_1,b_2,b_3)~~~
\ee
two triplets, the singlets contained in their product are given by
\be
\begin{array}{llll}
1&\equiv(ab)&=&(a_1 b_1+a_2 b_3+a_3 b_2)\\
1'&\equiv(ab)'&=&(a_3 b_3+a_1 b_2+a_2 b_1)\\
1''&\equiv(ab)''&=&(a_2 b_2+a_1 b_3+a_3 b_1)
\label{dec2}
\end{array}
\ee
The two triplets can be separated into a symmetric and an antisymmetric part:
\bea
3_S\equiv(ab)_S&=&\frac{1}{3}(2 a_1 b_1-a_2 b_3-a_3 b_2,2 a_3 b_3-a_1 b_2-a_2 b_1,2 a_2 b_2-a_1 b_3-a_3 b_1)\cr
3_A\equiv(ab)_A&=&\frac{1}{2}(a_2 b_3-a_3 b_2,a_1 b_2-a_2 b_1,a_3 b_1-a_1 b_3)
\label{dec3}
\eea
Moreover, if $c$, $c'$ and $c''$ are singlets transforming as $1$, $1'$ and $1''$, and
$a=(a_1,a_2,a_3)$ is a triplet, then the products $ac$, $ac'$ and $ac''$ are triplets
explicitly given by $(a_1 c,a_2 c, a_3 c)$, $(a_3 c',a_1 c', a_2 c')$ and $(a_2 c'',a_3 c'', a_1 c'')$,
respectively.
Note that due to the choice of complex representation matrices for the
real representation 3 the conjugate $a^{\star}$ of $a \sim 3$ does not
transform as $3$, but rather $(a_1^{\star}, a_3^{\star}, a_2^{\star})$
transforms as triplet under $A_4$.
The reason for this is that
$T^{\star}= U^{T} T U$ and
$S^{\star}=U^{T} S U =S$ where $U$ is the matrix which exchanges the 2nd and 3rd row and
column.

\vfill
\newpage
\appendix

\section*{B~~~Canonical Normalization of $K$ and $K^c$ \& Diagonalization of $y_l$}

We perform the following transformations on the fields present in the Lagrangian: at first $K$ and $K^{c}$
are brought into their canonical form and then $y_l$ is diagonalized.

To diagonalize the
hermitian matrices $K$ and $K^{c}$ we apply the unitary transformations $W$ and $W^{c}$:
\begin{equation}
W^{\dagger} K W = \mbox{diag} ~~~~~~~ \mbox{and} ~~~~~~~ (W^c)^{\dagger} K^c W^c = \mbox{diag}~~.
\end{equation}
Normalizing $K$ and $K^c$ requires a rescaling of the fields via the real (diagonal)
matrices $R$ and $R^c$:
\begin{equation}
R W^{\dagger} K W R = \mathbb{1} ~~~~~~~~~\mbox{and} ~~~~~~~~~R^c (W^c)^{\dagger} K^c W^c R^c = \mathbb{1}~~.
\end{equation}
The fields $l$ and $e^c$ present in the original Lagrangian (here $e^c$ stands for the three SU(2) singlet leptons) are expressed as:
\begin{equation}
l=W R \, l^{\prime} ~~~~~~~~~\mbox{and} ~~~~~~~~~ e^c = W^c R^c (e^c)^{\prime}
\end{equation}
so that $\bar{l} K \, l = \bar{l}^{\prime} [R W^{\dagger} K W R] \, l^{\prime} = \bar{l}^{\prime} \mathbb{1}
\, l^{\prime}$ and $\bar{e}^c K^c e^c = (\bar{e}^c)^{\prime} [R^c (W^c)^{\dagger} K^c W^c R^c] (e^c)^{\prime}
= (\bar{e}^c)^{\prime} \mathbb{1} (e^c)^{\prime}$ holds. $y_l$ and
$\mathcal{M}$ given in the basis $l^{\prime}$ and $(e^c)^{\prime}$ take the form:
\begin{equation}
(e^c)^{T} y_l \, l = (e^{c})^{\prime \, T} R^c (W^c)^{T} y_l W R \, l^{\prime}~~,
\end{equation}
and
\begin{equation}
(e^c)^{T} \mathcal{M} \, l = (e^{c})^{\prime \, T} R^c (W^c)^{T}
\mathcal{M} W R \, l^{\prime}~~.
\end{equation}
We diagonalize the resulting Yukawa couplings of the charged leptons, $R^c (W^c)^{T} y_l W R$, by the usual bi-unitary
transformation:
\begin{equation}
U^{T} [R^c (W^c)^{T} y_l W R] V = \frac{\sqrt{2}}{v}\, \mbox{diag} (m_e,m_\mu,m_\tau)
\end{equation}
and arrive at the mass eigenbasis $l^{\prime\prime}$ and $(e^c) ^{\prime\prime}$:
\begin{equation}
(e^c)^{\prime} = U (e^c)^{\prime\prime} ~~~~~~~~~\mbox{and} ~~~~~~~~~ l^{\prime} = V l^{\prime\prime}~~~.
\end{equation}
Finally, the matrix $\mathcal{M}$ for the dipole moments is given as:
\begin{equation}
(e^c) ^{T} \mathcal{M} \, l = (e^c) ^{\prime\prime \, T} [U^{T} R^c (W^c)^{T} \mathcal{M} W R
V] \, l^{\prime\prime} \equiv (e^c) ^{\prime\prime \, T} \mathcal{\hat{M}} \, l^{\prime\prime}~~~.
\end{equation}
Using $\mathcal{\hat M}$ we can immediately read off the size of MDMs, EDMs and
LFV processes.

As we assume for the actual calculation of $\mathcal{\hat M}$ that all couplings involved are
real, the matrices $W$, $W^c$, $U$ and $V$ turn out to be orthogonal instead of unitary. Furthermore,
we express the small parameter $u$ in terms of $t$ as $u = x t$ with $|x| \leq 1$ according to eq. (\ref{tbound}) and
eq. (\ref{ubound}). We then can do the calculation in just one expansion parameter $t$ \footnote{The different factors of $t$
and $u$ can be recovered in the final result by replacing $x$ with $\frac{u}{t}$.}.
In the course of the calculation we pose the following
requirements: the kinetic terms are canonically normalized up to and including $O(t^5)$,
$y_l$ is diagonal also up to and including $O(t^5)$, and the matrices $W$, $W^c$, $U$ and $V$
are orthogonal up to the same order. The calculations have been performed with two independent methods.

\end{document}